%% file: main.tex
\documentclass[acmsmall]{acmart}

\input{macro}

\begin{document}

\title[\resizebox{4.5in}{!}{When Fine-Tuning LLMs Meets Data Privacy: An Empirical Study of Federated Learning in LLM-Based Program Repair}]{When Fine-Tuning LLMs Meets Data Privacy: An Empirical Study of Federated Learning in LLM-Based Program Repair}
\input{authors}

\input{0.abstract}
\input{concepts}


\maketitle

\input{1.introduction}
\input{2.background}
\input{3.relatedwork} 
\input{4.studydesign}
\input{5.experiments}
\input{6.discussion}

\input{7.threats}
\input{8.conclusion}

\bibliographystyle{ACM-Reference-Format}
\bibliography{mybibs}

\end{document}

%% file: macro.tex
\usepackage{amsmath,amssymb,amsfonts}
\usepackage{graphicx}
\usepackage{textcomp}
\usepackage{xcolor}
\usepackage{tabularray}
\usepackage{svg}
\usepackage[normalem]{ulem}
\usepackage{tcolorbox}
\usepackage{lipsum}
\usepackage{longtable}
\usepackage{makecell}
\usepackage[]{tcolorbox}
\usepackage{hanging}
\usepackage{tcolorbox}
\usepackage{subcaption}
\usepackage{footnote}
\usepackage{url}
\usepackage{threeparttable}
\usepackage{comment}
\usepackage{adjustbox}
\usepackage{algorithmicx}
\usepackage{algpseudocode}
\usepackage{array}
\usepackage{balance}
\usepackage{booktabs}
\usepackage[skip=1pt,labelfont=bf]{caption}
\usepackage{calc}
\usepackage{calligra}
\usepackage{color}
\usepackage{colortbl}
\usepackage{courier}
\usepackage{csvsimple}
\usepackage{enumitem}
\usepackage{fancybox}
\usepackage{fontenc}
\usepackage{fontawesome5}
\usepackage{graphicx}
\usepackage{listings}
\usepackage{longtable}
\usepackage{lscape}
\usepackage{makecell}
\usepackage{marvosym}
\usepackage{moreverb}
\usepackage{multicol}
\usepackage{multirow}
\usepackage{pifont}
\usepackage{pgfplots}
\usepackage{rotating}
\usepackage{setspace}
\usepackage{siunitx}
\usepackage{soul}
\usepackage{subcaption}
\usepackage{tablefootnote}
\usepackage[]{tcolorbox}
\usepackage[normalem]{ulem}
\usepackage{url}
\usepackage{wasysym}
\usepackage{xspace}

\usepgfplotslibrary{statistics}
\usetikzlibrary{matrix, shapes.geometric, arrows}
\usetikzlibrary{shapes, arrows, positioning}

\algnewcommand\algorithmicforeach{\textbf{for each}}
\algdef{S}[FOR]{ForEach}[1]{\algorithmicforeach\ #1\ \algorithmicdo}

\newcolumntype{L}[1]{>{\raggedright\let\newline\\\arraybackslash\hspace{0pt}}m{#1}}
\newcolumntype{C}[1]{>{\centering\let\newline\\\arraybackslash\hspace{0pt}}m{#1}}
\newcolumntype{R}[1]{>{\raggedleft\let\newline\\\arraybackslash\hspace{0pt}}m{#1}}

\definecolor{codegreen}{rgb}{0,0.6,0}
\definecolor{codered}{rgb}{1,0,0}
\definecolor{codegray}{rgb}{0.5,0.5,0.5}
\definecolor{codepurple}{rgb}{0.58,0,0.82}
\definecolor{backcolour}{rgb}{0.95,0.95,0.92}
\definecolor{lightgray}{gray}{0.9}

\newboolean{showcomments}
\setboolean{showcomments}{false}
\ifthenelse{\boolean{showcomments}}
 { \newcommand{\mynote}[2]{
      \fbox{\bfseries\sffamily\scriptsize#1}
        {\small$\blacktriangleright$\textsf{\emph{#2}}$\blacktriangleleft$}}}
        { \newcommand{\mynote}[2]{}}
        
\definecolor{DarkOrange}{rgb}{0.8,0.3,0.0}
\definecolor{DarkCyan}{rgb}{0.0, 0.55, 0.55}
\definecolor{DarkCyel}{rgb}{1.0, 0.49, 0.0}
\definecolor{yellow-green}{rgb}{0.6, 0.8, 0.2}

\newcolumntype{?}{!{\vrule width 1pt}}

\newcommand{\todoc}[2]{{\textcolor{#1} {\textbf{#2}}}}

\newcommand{\todoblue}[1]{\todoc{blue}{\textbf{#1}}}
\newcommand{\todogreen}[1]{\todoc{yellow-green}{\textbf{#1}}}

\newcommand{\todocyel}[1]{\todoc{DarkCyel}{\textbf{#1}}}

\newcommand{\luo}[1]{\mynote{Luo}{\todocyel{#1}}}

\newcommand{\bachle}[1]{\mynote{Bach}{\todogreen{#1}}}
\newcommand{\ye}[1]{\mynote{Ye}{\todoblue{#1}}}

\lstdefinelanguage{mymarkdown}{
    morekeywords={*,\#, \#\#, \#\#\#},
    sensitive=false,
    morecomment=[l]{//},
    morestring=[b]",
    commentstyle=\color{codegreen},
    keywordstyle=\color{magenta},
    numberstyle=\tiny\color{codegray},
    stringstyle=\color{codepurple},
    basicstyle=\small,
    breakatwhitespace=false,         
    breaklines=true,
    breakindent=0pt,
    keepspaces=true,                 
    numbers=left,                    
    numbersep=5pt,                  
    showspaces=false,                
    showstringspaces=false,
    showtabs=false,                  
    tabsize=2,
}

\lstdefinestyle{mystyle}{
    commentstyle=\color{codegreen},
    keywordstyle=\color{magenta},
    numberstyle=\small\color{black},
    stringstyle=\color{codepurple},
    basicstyle=\scriptsize\ttfamily,
    breakatwhitespace=false,
    breaklines=true,
    captionpos=b,
    keepspaces=true,
    showspaces=false,
    showstringspaces=false,
    showtabs=false,
    tabsize=2
}

\lstdefinelanguage{diff}{
  morecomment=[f][\color{blue}]{@@},     
  morecomment=[f][\color{red}]-,         
  morecomment=[f][\color{codegreen}]+,       
  morecomment=[f][\color{red}]{---}, 
  morecomment=[f][\color{codegreen}]{+++},
  numberstyle=\tiny\color{codegray},
  numbers=left,                    
  numbersep=5pt,         
}

\setlist{noitemsep} 

\definecolor{darkpastelred}{rgb}{0.76, 0.23, 0.13}
\definecolor{ao(english)}{rgb}{0.0, 0.5, 0.0}

\definecolor{darkpastelred}{rgb}{0.76, 0.23, 0.13}
\definecolor{ao(english)}{rgb}{0.0, 0.5, 0.0}

\newboolean{useblue}
\setboolean{useblue}{false} 

\newcommand{\maybeblue}[1]{%
    \ifthenelse{\boolean{useblue}}%
    {\textcolor{blue}{#1}}%
    {#1}%
}

\AtBeginDocument{%
  }

\setcopyright{acmlicensed}
\copyrightyear{2024}
\acmYear{2024}
\acmDOI{XXXXXXX.XXXXXXX}


%% file: authors.tex

\author{Wenqiang Luo}
\affiliation{%
  \institution{Department of Computer Science, City University of Hong Kong}
  \country{China}
}
\email{wenqialuo4-c@my.cityu.edu.hk}

\author{Jacky Wai Keung}
\email{Jacky.Keung@cityu.edu.hk}
\affiliation{%
  \institution{Department of Computer Science, City University of Hong Kong}
  \country{China}
}

\author{Boyang Yang}
\affiliation{%
  \institution{Jisuan Institute of Technology, Beijing JudaoYouda Network Technology Co. Ltd.}
  \country{China}}
\email{yangboyang@jisuanke.com}

\author{He Ye}
\affiliation{%
  \institution{School of Computer Science, Carnegie Mellon University}
  \country{USA}}
\email{hey@cs.cmu.edu}

\author{Claire Le Goues}
\affiliation{%
  \institution{School of Computer Science, Carnegie Mellon University}
  \country{USA}}
\email{clegoues@cs.cmu.edu}

\author{Tegawend\'e F. Bissyand\'e}
\affiliation{%
  \institution{SnT, University of Luxembourg}
  \country{Luxembourg}}
\email{tegawende.bissyande@uni.lu}

\author{Haoye Tian}
\authornote{Corresponding author.}
\affiliation{%
  \institution{School of Computing and Information Systems, University of Melbourne}
  \country{Australia}
}
\email{tianhaoyemail@gmail.com}

\author{Bach Le}
\affiliation{%
  \institution{School of Computing and Information Systems, University of Melbourne}
  \country{Australia}
}
\email{bach.le@unimelb.edu.au}


\renewcommand{\shortauthors}{Luo et al.}

%% file: 0.abstract.tex
\begin{abstract}
  
\bachle{This paragraph should introduce LLMs like you did, and then introduce the fact that often LLMs are fine-tuned for a particular task at hand. Then say, but to fine-tune it introduces the inherent risks of data privacy. From there, motivate your study to address this problem.}\luo{\faCheckSquare[regular] Added the transitive description upon fine-tuning LLMs for specifc tasks.}
\ye{Simplify below to highlight the key problem of data privacy. The relationship between APR and data privacy is not clear. Say something about why  should APR concerned about data privacy?}\luo{\faCheckSquare[regular] Revised the motivation for APR, and how APR relates to privacy.}
Software systems have been evolving rapidly and inevitably introducing bugs at an increasing rate, leading to significant losses in resources consumed by software maintenance. Recently, large language models (LLMs) have demonstrated remarkable potential in enhancing software development and maintenance practices, particularly in automated program repair (APR) with improved accuracy and efficiency of bug fixing. However, LLM-based APR heavily relies on high-quality code repositories. A larger portion of existing code repositories are for private use and proprietary assets from various industries, reflecting more diversity and nuances in the data since real-world industries often have more extensive software development practices, which cannot be covered by merely public datasets. Therefore, utilizing private datasets shows significant potential in enhancing software development and maintenance. However, obtaining such data from various industries is hindered by data privacy concerns, as companies are reluctant to share their proprietary codebases. Nevertheless, there has also been no in-depth investigation of collaborative software development by learning from private and decentralized data while preserving data privacy for program repair in prior works.

\ye{Motivation is not clear. Why federated learning, why program repair? I will come back to this after I read the whole paper.}\luo{\faCheckSquare[regular] Revised the motivation for federated learning.}
To address the gap, we investigate the use of federated learning as a privacy-preserving approach that enables private entities to fine-tune LLMs on proprietary and decentralized data, facilitating the collaboration between clients to fully utilize their data to help enhance software development and maintenance. We conduct experiments using a private industrial dataset TutorCode for fine-tuning and the augmented benchmark EvalRepair-Java for evaluation. We examine whether and to what extent federated fine-tuning can enhance program repair. We construct heterogeneous code, which exhibits variations in code features such as coding style and code complexity across all clients, to explore the bug fixing capability of LLMs fine-tuned with different degrees of heterogeneity. We also assess the influence of federated learning algorithms optimized for different phases of federated learning on program repair to further investigate practical implications for real-world software development collaboration.
Our evaluation reveals that federated fine-tuning can effectively enhance program repair capabilities, with a maximal increase of 16.67\% and 18.44\% on $Top@10$ and $Pass@10$, respectively. It even exhibits bug fixing capabilities comparable to the centralized learning approach, which is regarded as the upper bound for traditional federated learning tasks since it is the ideal scenario where all data are aggregated together for learning. \bachle{Not clear what is meant here}\luo{\faCheckSquare[regular] Revised the explanation.} Notably, the impact of heterogeneous code on LLM fine-tuning is negligible, indicating that real-world industries can benefit from collaborative development regardless of diverse data distributions. Furthermore, each type of federated algorithm exhibits unique strengths across different LLMs, suggesting that fine-tuning for program repair can be enhanced by tailoring the optimization process to specific characteristics of different LLMs.


\end{abstract}

%% file: concepts.tex
\ccsdesc[500]{Software and its engineering~Software testing and debugging}

%% file: 1.introduction.tex
\section{Introduction}
Learning-based automated program repair (APR) has attracted considerable attention with emerging deep learning (DL) techniques, significantly reducing the manual effort in software maintenance \cite{zhang2023survey}.
The continuous evolution of modern software systems inevitably leads to the emergence of software bugs at an increasing rate, which results in substantial software maintenance costs in terms of both time and financial resources \cite{weiss2007long,boulder2019university,britton2013reversible}.  
Benefiting from the advancements of large language models (LLMs), there is a growing research interest in applying LLMs to the bug-fixing process of APR, which further improves the efficiency of automated bug repair. Fine-tuning LLMs to satisfy specific downstream tasks is rather effective since they have been pre-trained with huge corpora. Recent studies have demonstrated the potential of fine-tuning LLMs for APR tasks, showing significant improvements in repair accuracy and generalization \cite{jiang2023impact,huang2023empirical}. 
Novel approaches have been explored to fine-tune LLMs for APR, 
including standard neural machine translation (NMT) fine-tuning \cite{jiang2021cure,fu2022vulrepair}, multi-objective fine-tuning \cite{yang2024multi}, incorporating additional information such as code review to enhance fine-tuning \cite{paul2023enhancing}, augmenting fine-tuning datasets with syntactically diverse but semantically equivalent code \cite{hao2023enhancing}, utilizing optimal code representations \cite{silva2023repairllama} and parameter-efficient techniques are also adopted for effective and low-cost fine-tuning \cite{zirak2024improving,yang2024multi,silva2023repairllama}, each aiming to enhance program repair capabilities through LLM fine-tuning. \bachle{Citations to a few fine-tuned LLM-based APR here.}\luo{\faCheckSquare[regular] Added supportive related works regarding LLM fine-tuning for APR.}
\ye{Agreed with Bach. I added more works and categories.}

The access to extensive datasets has enabled LLMs to unleash their potential to learn complex patterns and nuances in different scenarios. Researchers for a long period have been mining public communities millions of code samples in the form of bug-fix pairs to enhance the repairing ability of their approaches \cite{tufano2019empirical, jin2023inferfix}. LLM-based APR is further improved by fine-tuning with carefully curated open-source data. Moreover, researchers and developers use synthetic data generated by models to enhance fine-tuning and satisfy specific needs \cite{kalyan2023survey,xia2022practical,wang2023codet5+}. By fine-tuning with a variety of data, the LLMs are capable of repairing bugs across different bug types, scenarios, and programming languages \cite{yuan2022circle, zhang2024appt, huang2023empirical}. The fine-tuning paradigm, to a large extent, has enabled the LLMs to achieve better performance and adapt to specific downstream tasks. 

\ye{In this paragraph, I feel like you should talk about why APR cares about data privacy. APR aims to modify program behavior, right? How does it relate to data privacy? }\luo{\faCheckSquare[regular] Revised the relation between APR and privacy.}
However, as a promising technique to enhance software maintenance, LLM-based APR heavily relies on high-quality code repositories since the effectiveness of LLM fine-tuning largely depends on the size and quality of the fine-tuning dataset \cite{zhang2024when,woisetschlager2024federated}.
\ye{The open source dataset is already huge enough.  I am not sure to mention 'Internet companies and technology companies that are not publicly available.'}\luo{\faCheckSquare[regular] Revised.}
While there are millions of public repositories on open-source platforms such as GitHub \cite{shanbhag2022exploring}, a larger proportion of existing code repositories are for private use and proprietary assets from various industries\footnote{\url{https://github.com/about}}\footnote{\url{https://web.archive.org/web/20240923210406/https://github.com/search?q=is\%3Apublic&type=repositories}}. Such proprietary industrial repositories often contain data reflecting a wealth of diversity and nuances since real-world industries employ various software development practices and handle different unexpected requirements \cite{lwakatare2020large}. Therefore, simply relying on public data is insufficient to build robust and effective software systems. However, direct collaboration between industries to make full use of private data is nearly impossible due to the problem of data privacy. On the other hand, the pre-training corpus of state-of-the-art LLMs, such as GPT-4 \cite{achiam2023gpt}, has a knowledge cut-off date of 2023, necessitating the continuous absorption of new data to facilitate their development and keep up-to-date with the latest advancements. Furthermore, despite the satisfactory performance achieved by existing academic research, it is challenging to transform them into practical use for industries since those academic models may underperform with unseen private industrial data \cite{yang2024federated}. Another critical problem pointed out by a group of researchers is that high-quality publicly available data is expected to be depleted by 2026 \cite{villalobos2024position}. Public data and synthetic data can gradually become insufficient to meet the needs of research and software development. Therefore, it is necessary and promising to make use of private industrial data to narrow the gap between academic findings and practical implementations while preserving data privacy. 

\bachle{The paragraph below is very good, but it still lacks one or a few key sentences that explain the key enablers of federated learning that allow learning from private data while maintaining privacy. What are the key technological advances of federated learning here?}\luo{\faCheckSquare[regular] Added major advances of federated learning in privacy preservation.}
Federated learning has emerged as a promising approach that facilitates private entities to utilize their data collaboratively, while addressing the concern of data privacy by learning a model without exposing the raw data of each client \cite{mcmahan2017communication}. By leveraging this privacy-preserving approach, real-world industries such as software companies can benefit from collaborative learning to help improve software development and maintenance. Besides, data privacy can be further preserved by major advances of federated learning such as differential privacy \cite{yang2023dynamic}, which adds noise to model updates to protect sensitive data, and secure aggregation protocols \cite{kairouz2021distributed} that ensure private updates are not revealed to the server. Additionally, secure multi-party computation \cite{xu2019hybridalpha} allows computations to be performed on encrypted data, further enhancing privacy assurance. By learning from a diversity of datasets from different devices, federated learning has the potential to continuously improve model performance beyond privacy preservation. In particular, federated learning has also exhibited its capability of improving software development. For example, a federated learning framework ALMITY \cite{yang2024federated} has been proposed to enhance model training for code clone detection and defect prediction while tackling data imbalance. Integrating systematic debugging with a federated learning framework has also been found to be an effective fault localization approach \cite{gill2023feddebug}. Effective collaborative learning can be enabled by federated learning to achieve comparable performance with centralized learning on commit classification and code summarization tasks \cite{shanbhag2022exploring,kumar2024code}. Moreover, a few of the previous works have demonstrated the potential of enhancing LLM-based fine-tuning with federated learning while protecting data privacy \cite{chen2023federated,wang2023can,liu2023differentially}. LLM-based federated learning has been successfully applied to different other domains including knowledge distillation \cite{ma2023fedid}, recommendation \cite{zhao2024llm}, legal intelligence \cite{yue2023fedjudge} and model security \cite{zheng2023input}. A few fine-tuning strategies have also been proposed to mitigate the cost of communicating huge amounts of LLM parameters \cite{che2023federated,sun2023fedbpt}. 


\textbf{This work.} To address the gap in collaborative and decentralized software development, we conduct a comprehensive empirical study on fine-tuning LLMs for program repair using federated learning. In particular, we answer the following three research questions: RQ1 examines whether and to what extent can federated fine-tuning enhance program repair capabilities. RQ2 investigates the impact of fine-tuning LLMs with heterogeneous code in federated learning. RQ3 explores the influence of federated algorithms that enable optimization in different phases of federated learning on fine-tuning LLMs for program repair. We fine-tune various elaborately selected LLMs of code with a private industrial dataset that consists of 1239 real-world private programs and eliminates the data leakage problem. The fine-tuned LLMs are assessed on a benchmark substantially enhanced with up to 583 test cases for each problem to mitigate the issue of patch overfitting and perform a comprehensive and rigorous evaluation. Specifically, we conduct a series of experiments and analyses for each RQ as follows:

(1) Firstly, our study investigates the feasibility and effectiveness of enhancing LLMs while preserving data privacy in a decentralized and collaborative software development environment. We compare the performance of federated fine-tuning with standard fine-tuning methods including centralized learning, which is the ideal scenario that pools all data together, and the original LLMs across various model architectures and metrics to validate whether LLM-based federated fine-tuning is able to enhance program repair. We find that federated fine-tuning can substantially enhance the program repair capabilities of LLMs to achieve performance even comparable to the centralized fine-tuning approach, achieving maximal increases of up to 16.57\% on $Top@10$ and 18.44\% on $Pass@10$. CodeQWen-7B and CodeLlama-13B can fix 7 and 2 unique bugs on the augmented program repair benchmark, respectively, which are the most among all LLMs.

(2) Secondly, we explore the impact of data heterogeneity on the repairing capabilities of LLMs by constructing different degrees of heterogeneous data distributions based on various code features such as coding style, code complexity and code embedding to better align with real-world scenarios. Specifically, we extract 23 different attributes of coding style to identify and construct heterogeneous coding style. We establish heterogeneous code complexity based on the modified hunks between the buggy code and fixed code. Moreover, we construct heterogeneous code embedding by extracting the context information from the pairs of natural language and programming language. We evaluate the performance of LLMs fine-tuned with different data distributions to investigate how data heterogeneity influences program repair. Notably, the results indicate that heterogeneous code has no significant impact on the fine-tuning of LLMs compared to the ideal homogeneous data distribution. On the contrary, it leads to a considerable improvement that achieves maximal increases of 18.41\% and 21.46\% on $Top@10$ and $Pass@10$ for program repair. 

(3) Finally, we evaluate different federated learning algorithms optimized for various key stages of federated learning to explore their influence on LLM-based federated fine-tuning for program repair. We investigate different federated algorithms that are optimized for particular processes of federated learning, including client-side and server-side optimization. We also evaluate the personalized learning paradigm to assess its effectiveness in adapting LLMs to specific clients. Our analysis reveals that while different types of federated algorithms exhibit varying strengths and weaknesses across different LLMs, FedAvg demonstrates the best overall performance and the personalized learning approach remains a challenge for LLM fine-tuning.



\bachle{So far it's very good about the story that you wrote. In the above paragraph, you should present some key findings and experimental results. In the contribution summary below, summarize the key results too.}\luo{\faCheckSquare[regular] Elaborated the key insights and results in This work and Contributions.}

\textbf{Contributions.} In summary, our study makes the following main contributions:

\begin{itemize}
    \item We present an empirical study on fine-tuning LLMs for program repair with federated learning. Our study explores the feasibility and effectiveness of fine-tuning LLMs while {\bf preserving data privacy} in decentralized and collaborative software development. Our results indicate that federated fine-tuning significantly enhances program repair capabilities and achieves performance comparable to the regular centralized learning approach.
    \item We explore the influence of federated fine-tuning on program repair where the objective of such task is to generate correctly fixed patches, in contrast to the majority of prior federated learning research that concentrates on typical discriminative tasks, unveiling novel insights into the application of federated learning to other forms of generative tasks. It is worth noting that our results on program repair reveal the fact that the generative tasks may differ significantly compared to other types of tasks such as classification, regression, etc., in the context of decentralized LLM fine-tuning.
    \item We investigate the impact of different LLMs on the performance of federated program repair. By comparing various state-of-the-art LLMs of code that have varying architectures and pre-training strategies on multiple baselines, we provide insights into the suitability and practicality of different LLMs in federated learning. 
    \item We explore the effect of heterogeneous code on the program repair capability of LLMs in federated fine-tuning. We construct feature-skewed Non-IID (see Section \ref{dh}) data distribution to align with real-world scenarios where the collaborators have different code repositories with diverse software development practices and bug-fixing patterns. The results demonstrate that fine-tuning with different degrees of heterogeneous code causes negligible impact on the performance of LLMs and can still benefit program repair significantly. By evaluating the LLMs with different data distributions, we shed light on the robustness and adaptability of LLMs in handling Non-IID data in decentralized environments.
    \item We evaluate the impact of various federated learning algorithms that are optimized for different stages of federated learning on the bug fixing capability of LLM-based program repair. The results show that different optimizations exhibit varying strengths and weaknesses across different LLMs. Our analysis provides insights into the trade-offs and considerations when selecting federated learning algorithms for LLM-based program repair.

\end{itemize}

\textbf{Availability.} The artifact of this study is publicly available at: \textbf{\url{https://github.com/stringing/Federated-LLM-Based-APR}}.

The remainder of this paper is organized as follows: Section \ref{background} describes the background knowledge of this study. Section \ref{relatedwork} provides an overview of related work. Section \ref{studydesign} presents the study design, including the research questions and the overall methodology. Section \ref{experiment} details the experiments conducted and analyze the results obtained. Section \ref{disc} discusses additional considerations and aspects of this study. Potential threats to validity are identified in Section \ref{threats}. Finally, Section \ref{conclusion} concludes the paper by summarizing the main contributions and insights of this paper.

%% file: 2.background.tex
\section{Background}\label{background}

\subsection{Parameter-Efficient Federated Fine-tuning}\label{ff}
Federated learning is a distributed machine learning paradigm that enables multiple participants to collaboratively train a model without sharing their private data, thereby preserving privacy. Different from traditional distributed learning, where data collected from multiple sources are transferred to a server, and all data are placed together to train a centralized model, each device in federated learning trains a single model on its local dataset, and a central server is responsible for aggregating all models into a final global model.  

\begin{figure}[htbp]
    \centering
    \includegraphics[width=\textwidth, bb=0 0 770 284]{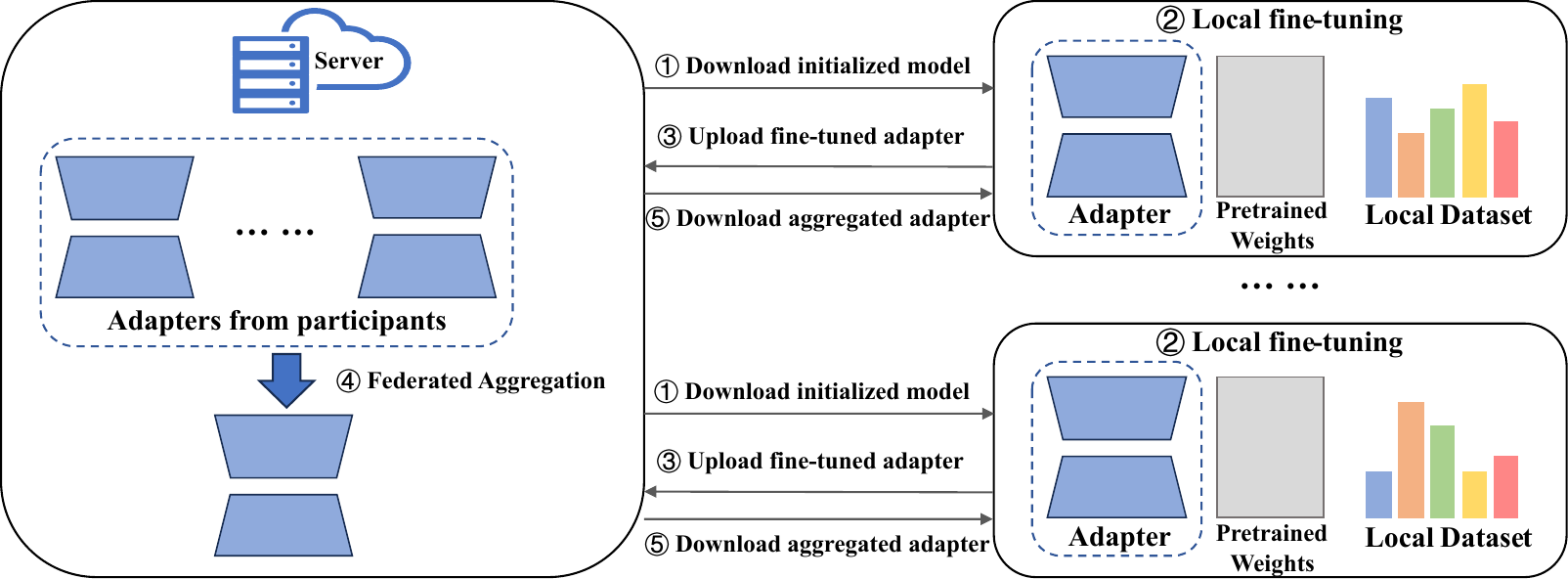}
    \caption{The workflow of federated learning of parameter-efficient federated fine-tuning framework.}
    \label{fl}
\end{figure}

Figure \ref{fl} presents the workflow of the parameter-efficient federated fine-tuning process. All clients are dispatched with the same randomly initialized model at first, and each participant then starts to fine-tune its model with its local dataset. Typically, the fine-tuned models are uploaded to the central server for further processing. However, the communication cost can be extremely high for LLMs since billions of parameters need to be transmitted. Therefore, benefit from lightweight fine-tuning techniques such as Low-Rank Adaption (LoRA) \cite{hu2021lora} and Quantized Low-Rank Adaption (QLoRA) \cite{dettmers2024qlora}, only the adapter that is a small fraction of parameters is supposed to be uploaded while the original pre-trained model weights are retained in the client's device. Once the central server has received all models from the clients, these models are aggregated into one global model, which is then dispatched to all clients for the next round. The global model attained in the final round is the final model that can be applied to downstream tasks. Note that the central server in federated learning acts as a trusted third-party server. Throughout the entire federated learning process, all datasets are kept local and not shared with any other clients, thereby ensuring data privacy is protected.

In our study, we utilize the efficient fine-tuning approach QLoRA, which is an extended version of LoRA in order to alleviate resource overhead in terms of GPU memory and parameter transfer in federated learning. The work of QLoRA proposed a novel 4-bit normal float (NF4) data type to quantize the pre-trained weights and reduce additional storage cost of quantization constants by double quantization \cite{dettmers2024qlora}. 

In detail, the workflow mainly consists of the following steps:

\noindent\textbf{(1) Model Initialization:} Initially, the pre-trained model weights $\boldsymbol{W_0}\in \mathbb{R}^{d_1\times d_2}$ on each client are quantized to 4-bit precision using the NF4 quantization format:
\begin{equation}
    \boldsymbol{\Tilde{W_0}}=Q_{NF4}(\boldsymbol{W_0})
\end{equation}
where $Q_{NF4}$ denotes the quantization function and $\boldsymbol{\Tilde{W_0}}$ represents the quantized pre-trained model weights. The quantized weights $\boldsymbol{\Tilde{W_0}}$ remain frozen during fine-tuning while the trainable adapter $\boldsymbol{\Delta W}=\boldsymbol{BA}$, where $\boldsymbol{B}\in \mathbb{R}^{d_1\times r}$ and $\boldsymbol{A}\in \mathbb{R}^{r\times d_2}$ are the low-rank decomposed matrices, is introduced into the model architecture. $r\ll min(d_1,d_2)$ is the rank of adaption and also a hyperparameter in our setting. In order to preserve model consistency in federated fine-tuning, the random Gaussian initialization for $\boldsymbol{A}$ is performed on the server and then it is downloaded by all clients as the initial parameters at the beginning of fine-tuning, except for $B$ whose initial weights are zero.

\noindent\textbf{(2) Local Fine-tuning:} Consider a federated learning system composed of $K$ clients, each holding a local fine-tuning dataset $\mathcal{D}_k$, where $k\in {1,2,...,K}$. The adapted model $\boldsymbol{W_k}$ at client $k$ is:
\begin{equation}
    \boldsymbol{W_k}=\boldsymbol{\Tilde{W_0}}+\boldsymbol{\Delta W_k}=\boldsymbol{\Tilde{W_0}}+\boldsymbol{B_kA_k}
\end{equation}
The fine-tuning objective is to minimize the local loss function $\mathcal{L}_k$ with respect to the adapter parameters $\boldsymbol{B_k}$ and $\boldsymbol{A_k}$:
\begin{equation}
    \mathcal{L}_k\left( \boldsymbol{\Tilde{W_0}}+\boldsymbol{B_k}\boldsymbol{A_k};\mathcal{D}_k \right)=\frac{1}{n_k}\sum_{i=1}^{n_k}\ell\left( f\left( \boldsymbol{\Tilde{W_0}}+\boldsymbol{B_k}\boldsymbol{A_k};\boldsymbol{x_i^k} \right),\boldsymbol{y_i^k} \right)
\end{equation}
where $(\boldsymbol{x_i^k},\boldsymbol{y_i^k})\in \mathcal{D}_k$ is one of the bug-fix code pairs in the fine-tuning dataset $\mathcal{D}_k$ and $n_k$ is the size of $\mathcal{D}_k$. $f$ is the output function of the model and we employ a cross-entropy loss function $\ell$ to measure the difference between the probability distributions of the model output and actual code for each token in the sequences as follows:
\begin{equation}
    \ell(\boldsymbol{\hat{y}_i^k},\boldsymbol{y_i^k};\boldsymbol{x_i^k})=-\sum_j Q\left( \hat{y}_{i,j}^k\Big| \boldsymbol{x_i^k},\hat{y}_{i,1}^k,...,\hat{y}_{i,j-1}^k \right)\times \log P\left( y_{i,j}^k\Big| \boldsymbol{x_i^k},y_{i,1}^k,...,y_{i,j-1}^k \right)
\end{equation}
where $Q$ is the probability distribution for the predicted output $\boldsymbol{\hat{y}_i^k}$ and $P$ is the actual distribution for $\boldsymbol{y_i^k}$. $\hat{y}_{i,j}^k\in\boldsymbol{\hat{y}_i^k}$ and $y_{i,j}^k\in\boldsymbol{y_i^k}$ are one of the tokens in the predicted output and actual code respectively. Then the optimized low-rank adapter parameters can be obtained by using various optimizers such as SGD:
\begin{equation}
    \boldsymbol{\Delta W_k^*}=\boldsymbol{B_k^*}\boldsymbol{A_k^*}=\boldsymbol{\Delta W_k}-\eta\Delta\mathcal{L}_k\left( \boldsymbol{\Tilde{W_0}}+\boldsymbol{B_k}\boldsymbol{A_k};\mathcal{D}_k \right)
\end{equation}
where $\eta$ is the learning rate. Note that in this framework the paged optimizers are utilized according to the setting of QLoRA, which are able to automatically transfer the information from GPU to CPU for error-free GPU processing when GPU encounters out-of-memory issues \cite{dettmers2024qlora}.

\noindent\textbf{(3) Fine-tuned Model Upload:} Instead of uploading the entire fine-tuned LLMs, clients upload only the fine-tuned adapters $\boldsymbol{B_k^*}\boldsymbol{A_k^*}$ to the central server. This significantly reduces the cost of transferring model parameters since the adapter constitutes only a small portion of the original model parameters.

\noindent\textbf{(4) Federated Aggregation:} In this step, the central server aggregates the adapter parameters received from all clients into a global adapter $\boldsymbol{\Delta W_G}$ using aggregation algorithms such as federated averaging (FedAvg) \cite{mcmahan2017communication}:
\begin{equation}
    \boldsymbol{\Delta W_g}=\boldsymbol{B_g}\boldsymbol{A_g}=\frac{1}{n}\sum_{k=1}^Kn_k\boldsymbol{B_k^*}\boldsymbol{A_k^*}
\end{equation}
where $n=\sum_{k=1}^Kn_k$.

\noindent\textbf{(5) Aggregated Model Download:} The aggregated adapter $\boldsymbol{\Delta W_g}$ is dispatched back to all clients. Each client updates the local adapter with the downloaded global adapter:
\begin{equation}
    \boldsymbol{\Delta W_k^\prime}=\boldsymbol{\Tilde{W_0}}+\boldsymbol{\Delta W_g}
\end{equation}
The updated model $W_k^\prime$ serves as the starting point for the next round of federated fine-tuning. The global model obtained after a total of $T$ rounds of fine-tuning is the final model, which can be further applied to corresponding downstream tasks.

Throughout the entire federated fine-tuning process, parameter efficiency is achieved by quantization, significantly reducing storage costs for each client and the overhead of parameter transfer in federated learning system. Data privacy is also protected by not exposing local data to each participating client. In this study, we investigate whether effective fine-tuning can be promoted by this framework to achieve comparable performance while protecting data privacy at the same time.

\subsection{Data Heterogeneity in Federated Learning}\label{dh}
Data heterogeneity has been a critical challenge that incurs performance degradation in federated learning \cite{kairouz2021advances}. In ideal situations where the data across all clients are independently and identically distributed (IID), federated learning can achieve comparable performance without centralizing the data \cite{mcmahan2017communication}. However, real-world data distributions can be affected by various reasons. The datasets collected on each device tend to vary in their distributions due to a number of factors such as geographical differences, demographic differences and different user behaviors, which exhibit non-independently and identically distributed (Non-IID) characteristics. Therefore, each local model trains towards different objectives inconsistent with the global optimum and eventually deviate from the optimal solution under Non-IID scenarios \cite{li2022federated}. According to the works of Kairouz et al. \cite{kairouz2021advances}, there are mainly three basic types of Non-IID data. Consider the local data distribution $P(\boldsymbol{x_i^k},\boldsymbol{y_i^k})$ at client $k$, the three fundamental Non-IID scenarios are as follows:

\noindent\textbf{(1) Label skew.} Label skew arises when the distribution of the target labels differs across devices, i.e., the marginal distribution $P(\boldsymbol{y_i^k})$ varies in each client. Label skew happens when certain labels are more prevalent on some devices than others, leading to an imbalance in the label distribution. For example, rice is commonly grown in Southeast Asia, whereas in drier areas like North America, such type of crop is mainly wheat. Generally, such imbalances can significantly affect the performance of DL models in federated learning, which may become biased towards the more prevalent classes. However, different from typical supervised tasks where the data is accompanied with fixed and explicit labels, we do not focus on label skew in this study for program repair task where the objective is to generate a valid patch that correctly fixes a bug.

\noindent\textbf{(2) Feature skew.} Feature skew occurs when the distribution of the input features varies across different devices, i.e., $P(\boldsymbol{x_i^k})$ differs across clients. For example, the same handwritten number from different people can present in different ways because of diverse writing habits. In the context of software engineering, feature skew can manifest in various ways due to the diverse nature of programming languages, frameworks and development practices. For instance, consider a scenario where client $A$ primarily develops applications using Java and the Spring framework, while Client $B$ focuses on Python and the Django framework. The code features, such as syntax, libraries and design patterns, will exhibit significant differences and variations in the feature representation between these two clients. Consequently, the local models trained on these distinct feature distributions may struggle to generalize effectively to the global model, which can lead to suboptimal performance in federated learning. We construct and explore various code features in this study to investigate the impact of heterogeneous code on federated learning.

\noindent\textbf{(3) Quantity skew.} Quantity skew (i.e., $n_k$ varies across clients) exists when different clients hold different amounts of data. In real-world scenarios, companies of varying sizes or from different industries often possess diverse quantities of data due to differences in business scale, data collection capabilities, etc. This disparity in data quantity can lead to an imbalance in the contributions of participating devices during federated learning, potentially introducing a bias in the global model. On the other hand, label-skewed or feature-skewed data is often inherently accompanied with quantity skew. Quantity skew typically does not exist in isolation for most scenarios, but coexists with the other types of skews instead. Thus, we mainly focus on feature skew to investigate the variations in code rather than quantity in our study.

Overall, each type of skew is able to cause performance loss in federated learning since each model trained on different distribution has its own bias against data, which leads to instability and inconsistency in the global model aggregated from these individual models. Furthermore, whether the final model aggregated from models fine-tuned with diverse data distributions is negatively affected and to what extent it is affected still remains unclear in terms of LLM fine-tuning, especially with the generative tasks that mainly involve feature skew.

Therefore, as program repair, which aims to generate correctly repaired code patches, lacks fixed and explicit labels and present diverse features in the code, we focus on feature skew in our study.

%% file: 3.relatedwork.tex
\section{Related Work}\label{relatedwork}
\subsection{Federated Learning for Software Engineering}
\noindent\textbf{Existing Works.} Despite the growing interest in federated learning, its application to software engineering tasks remains relatively unexplored, with only a limited number of studies addressing this area. Yang et al. \cite{yang2024federated} proposed a novel federated learning framework ALMITY to enhance the models trained on private and heterogeneous datasets for code clone detection and defect prediction. ALMITY provided an approach to tackle data imbalance and improve minority class learnability. Gill et al. \cite{gill2023feddebug} integrated systematic debugging with federated learning to locate problematic clients with a novel replay strategy. In order to explore the value of non-open source projects for federated learning on software engineering,  Shanbhag et al. \cite{shanbhag2022exploring} conducted a study on commit classification to examine whether federated learning achieved comparable performance to centralized learning. Kumar et al. \cite{kumar2024code} evaluated federated learning on code summarization and found that federated learning is as effective as centralized learning.

\noindent\textbf{Limitations.} Existing studies suffer from several limitations. Most of them lack further investigation into LLMs with only a limited range of architectures or sizes, also failing to explore the potential of state-of-the-art models. Additionally, there is a lack of exploration into data heterogeneity, which limits understanding of the impact of diverse data during software development. In the meanwhile, the focus has predominantly been on tasks with labeled data, neglecting the crucial challenges posed by generation tasks with feature-skewed data, which is prevalent in real-world applications. Lastly, the reliance on simulated data derived from public datasets, rather than incorporating real-world private industrial data, limits the generalizability and practical relevance of the findings. 

\noindent\textbf{Addressing the Gap.} In this study, we evaluate the efficacy of federated learning for program repair using multiple state-of-the-art LLMs to investigate their fixing ability. We evaluate the impact of data heterogeneity by employing a private industrial dataset to construct Non-IID scenarios. As program repair is a code generation task without fixed and explicit labels, we exploit the rich feature information embedded within the code. Consequently, we extract diverse code features to conduct a thorough analysis of the influence of heterogeneous code on federated learning.

\subsection{Federated Learning with LLMs for Privacy Preservation}
\noindent\textbf{Existing Works.} Since federated learning enables decentralized fine-tuning of LLMs without exposing raw data, it opens up possibilities for collaborative learning while maintaining data privacy. The use of private industrial data for LLM fine-tuning while preserving data privacy is a critical challenge and is of great significance. Considering the limited availability of public domain data and the need to protect private data, Chen et al. \cite{chen2023federated} conducted an analysis between different components of federated learning with LLMs, which include federated pre-training, federated fine-tuning and federated prompt engineering, to explore solutions to potential issues that might occur when applying federated learning to LLMs. In order to further promote the privacy-preserving capabilities of federated learning, Wang et al. \cite{wang2023can} selected public data that has similar distribution to private data for LLM training and used LLMs as the teaching models in knowledge distillation to facilitate the training of private models on each device. Liu et al. \cite{liu2023differentially} specifically designed a privacy-preserving approach for federated learning by employing Gaussian noise to the updates of LLMs. 

Furthermore, some researchers apply federated learning with LLMs to various tasks since centralized training of LLMs fails to tackle privacy concerns. Zhao et al. \cite{zhao2024llm} performed federated fine-tuning on users' behavior data for federated recommendation based on LLMs. Yue et al. \cite{yue2023fedjudge} utilized a continual learning approach with federated learning for legal tasks on an LLM of size 7B. Zheng et al. \cite{zheng2023input} studied the attack performance of LLMs with federated learning. Given that there were few works on federated distillation with large-scale models, Ma et al. \cite{ma2023fedid} conducted federated distillation with LLMs for both homogeneous and heterogeneous settings.

On the other hand, one of the major challenges of federated learning with LLMs is the huge number of parameters each update of an LLM consists of. To mitigate this issue, Che et al. \cite{che2023federated} designed an approach that selected important layers of prompts for efficient prompt-tuning. Sun et al. \cite{sun2023fedbpt} proposed an effective federated learning framework that enables prompt exchanging between servers and clients rather than parameters to improve the performance of LLMs. However, the performance of prompt-tuning can be suboptimal since the prompts require elaborate design with prior knowledge \cite{gao-etal-2021-making}.

\noindent\textbf{Limitations.} Despite recent advancements on LLMs, there remains a significant research gap regarding the application of federated learning with LLMs to software engineering tasks, especially those pertaining to code. Given the inherent value of code repositories to enterprises, ensuring data privacy becomes a crucial concern. Current research does not sufficiently address how federated learning can be effectively utilized for code-related tasks while ensuring data privacy.

\noindent\textbf{Addressing the Gap.} Our focus lies on program repair, which is a crucial code-related task for software maintenance, by leveraging federated learning with LLMs. We aim to preserve the privacy of sensitive code repositories while effectively enhancing program repair capabilities.

\subsection{Fine-tuning LLMs for Program Repair}
\noindent\textbf{Existing Works.} Pre-trained on extensive corpora, LLMs can already achieve remarkable performance without additional training. For instance, the AlphaRepair framework, introduced by Xia et al. \cite{xia2022less}, employs a cloze-style technique to generate code patches leveraging the original CodeBERT \cite{feng2020codebert} without fine-tuning, presenting superior performance compared to many state-of-the-art APR techniques. Furthermore, Xia et al. \cite{xia2023automated} performed an extensive investigation into the direct application of various LLM configurations for APR. Their findings suggested that LLMs offer considerable potential for enhancing APR effectiveness. 

Given that the benefits of LLMs remain underexplored, there is a growing trend toward further enhancing LLMs with fine-tuning techniques. To validate the effectiveness of fine-tuning LLMs for APR, Jiang et al. \cite{jiang2023impact} conducted the first study that specifically evaluated code LLMs on multiple APR benchmarks. This study found that fine-tuned LLMs were able to significantly outperform existing DL-based APR techniques, demonstrating great potential in LLM-based APR. To further evaluate the fine-tuned LLMs, Huang et al. \cite{huang2023empirical} considered multiple LLM architectures, programming languages for a comprehensive investigation covering bugs, vulnerabilities and errors on 7 benchmarks, revealing that fine-tuned LLMs can significantly surpass the performance of previous state-of-the-art APR tools. 

Moreover, building upon the promising results of fine-tuning LLMs for program repair, researchers have proposed various techniques and approaches to further enhance the performance of LLM-based APR for specific needs. Mashhadi et al. \cite{mashhadi2021applying} made a preliminary attempt in fine-tuning CodeBERT \cite{feng2020codebert} to fix simple bugs in Java. Silva et al. \cite{silva2023repairllama} fine-tuned LLMs with various code representations to improve the fine-tuning performance, highlighting the importance of domain-specific and expert code representations for APR. On the other hand, integrating the power of LLMs with existing tools can be another promising direction for enhancing LLM-based APR performance. Jin et al. \cite{jin2023inferfix} proposed a novel APR approach, InferFix, which utilized a static analyzer to detect bugs and fine-tuned an LLM to produce appropriate patches for the bugs. Wei et al. \cite{wei2023copiloting} combined a completion engine with the fine-tuned LLMs to mitigate the limitations of LLMs in generating syntactically and semantically correct codes in order to generate more valid fixes.

With further exploration, another remarkable capability of LLMs in APR has exhibited in cross-language repairing. The study conducted by Ahmed et al. \cite{ahmed2022multilingual} demonstrated that LLMs had substantial capacity to achieve multi-lingual program repair. This study revealed that LLMs fine-tuned across multiple programming languages are more likely to improve the repairing performance as opposed to a single programming language. Yuan et al. \cite{yuan2022circle} proposed an APR framework CIRCLE to achieve multi-lingual program repair and address the need of increasing software requirements. CIRCLE fine-tuned LLMs in a continual learning manner to learn from multiple programming languages with a replay approach that mitigated the risk of forgetting previous languages to enable accurate and stable multi-lingual program repair. 

\noindent\textbf{Limitations.} Current explorations of LLM-based program repair techniques have primarily focused on centralized environments, and program repair techniques in distributed environments require further investigation. Developers often work together on software development, especially for large-scale projects, collaborating across different locations and time zones. This collaborative nature of software development presents unique challenges and opportunities for decentralized program repair techniques.

\noindent\textbf{Addressing the Gap.} We focus on LLM-based program repair with a decentralized setting to explore the potential of collaborative learning while preserving data privacy. We simulate a number of distributed devices and datasets in the federated learning system to investigate the practicality of federated fine-tuning and provide insights that may benefit further research for decentralized program repair or other code-related tasks.

%% file: 4.studydesign.tex
\section{Study Design}\label{studydesign}

\subsection{Research Questions}
\noindent\textbf{RQ1: Can federated learning enhance LLM fine-tuning in repairing programs while protecting data privacy?} While federated learning preserves data privacy of each client, whether it can improve and to what extent it can enhance the bug fixing capability of the LLMs still remains underexplored for most code-related tasks. As a preliminary RQ, we want to understand the feasibility and effectiveness of federated fine-tuning. Therefore, we compare federated fine-tuning with other standard fine-tuning approaches across different LLMs to examine its potential to enhance program repair.

\noindent\textbf{RQ2: How do data distributions in federated learning environments affect the repairing capabilities of LLMs?} Software development practices vary significantly across real-world industries. The code repositories held by different organizations, such as software companies, can exhibit heterogeneity in code features such as coding style, code complexity, etc., whereas data heterogeneity has hindered further improvement for most traditional federated learning tasks. Therefore, in this RQ, we aim to explore the influence of heterogeneous code by constructing different degrees of Non-IID scenarios to fine-tune and evaluate the LLMs. 

\noindent\textbf{RQ3: How do different federated learning algorithms perform in fine-tuning LLMs for program repair?} As key components of federated learning, the optimization strategies in different phases, including local training and federated aggregation, are crucial to enhance the federated learning process and adapt to various federated learning scenarios. This RQ aims to investigate how federated algorithms with different types of optimization influence the performance of federated fine-tuning. We evaluate federated algorithms covering client-side optimization, server-side optimization, and personalized learning to validate the impact of algorithms on program repair.

\subsection{Fine-tuning Dataset and Evaluation Benchmark}
\noindent\textbf{Selection Criteria.} To effectively select the fine-tuning dataset and evaluation benchmark, we take several critical criteria into consideration:
\begin{itemize}
    \item[--] Firstly, the overlap of pre-training data and evaluation benchmark can cause data leakage \cite{huang2023empirical}, which hinders the evaluation to reflect the real capability of the models since they tend to achieve excellent performance on data they have been trained in advance \cite{tian2022change,wu2024condefects}. On the other hand, patch overfitting \cite{ye2021automated,le2019reliability} is another major concern for the evaluation benchmark due to the limitations of the test cases. The generated patches can be incorrect even though they have passed the test cases in the benchmark if deficient test suites fail to validate key intentions in the code \cite{lin2022context,wang2020automated}. Therefore, preventing data leakage and minimizing patch overfitting is paramount. 
    \item[--] Secondly, the use of prevalent programming languages ensures compatibility and ease of integration with tools commonly employed in both academia and industry. To that end, according to the findings by Zhang et al. \cite{zhang2024systematic} and the trends in the use of programming languages on GitHub\footnote{\url{https://madnight.github.io/githut/\#/pull\_requests/2024/1}}, the programming languages widely adopted in current communities include Python, Java, C++, C and JavaScript. Therefore, we aim to choose datasets with programming languages that fall within this range.
    \item[--] Thirdly, employing private industrial datasets in terms of LLM fine-tuning provides a practical source of real-world data, enhancing the model's applicability and robustness. Additionally, high-quality test cases are essential for the evaluation benchmark, offering a comprehensive assessment of model performance across diverse scenarios. 
    \item[--] Furthermore, providing detailed metadata with the datasets significantly improves the fine-tuning process by offering valuable information for data preprocessing and the models to fine-tune with. 
\end{itemize}

Consequently, we select datasets for fine-tuning and evaluation based on these criteria collectively to facilitate effective model development and accurate evaluation.

\begin{itemize}
\item[--] \textbf{Fine-tuning Dataset.} In order to investigate the performance of federated fine-tuning with LLMs using private industrial data in real-world scenarios, we employ a private industrial dataset TutorCode \cite{yang2024cref} as the fine-tuning dataset in our study to satisfy real-world conditions to validate the effectiveness of federated fine-tuning. TutorCode is a curated collection of programming submissions sourced from the online programming education platform by 20 experts, ensuring integrity and quality by manual verification. To prevent data leakage, the dataset is not crawled from public repositories but collected from the company's proprietary data. The usage license further safeguards its integrity and confidentiality by preventing potential unauthorized crawling. The TutorCode dataset comprises 1,239 buggy C++ programs developed by 427 programmers, addressing 35 distinct programming problems. The dataset covers a wide range of algorithms, 
which cater to both novice and advanced levels, resulting in 12 tiers of problem difficulty. It includes a variety of bugs spanning multiple functions, while the corresponding fixes also cover from one to more than ten modified code hunks. Furthermore, each buggy code is accompanied with extra metadata, such as problem description, to provide helpful information to improve the repairing process.
Since buggy code originates from various programmers with different coding habits and programming skills, rich information lies within the code itself. By mining different code features, we can construct datasets of distinct distributions to further explore the impact of heterogeneous code on federated fine-tuning.

\item[--] \textbf{Evaluation Benchmark.} Despite that existing benchmarks such as Defects4J \cite{just2014defects4j} and ManyBugs \cite{le2015manybugs} have been widely used to evaluate the capability of bug fixing, they still exhibit certain limitations, including data leakage \cite{zhang2024systematic} and patch overfitting \cite{zhang2023survey} issues. Therefore, we mainly aim to select high-quality evaluation benchmark that satisfy the above criteria and mitigate potential risks as much as possible. We adopt EvalRepair-Java \cite{yang2024multi} as the evaluation benchmark on federated fine-tuning. EvalRepair-Java consists of 163 code samples and is built on top of the foundation of HumanEval-Java \cite{zheng2023codegeex}, which was widely used and specifically designed to avoid data leakage by excluding itself from the pre-training datasets of existing LLMs. EvalRepair-Java provides enhanced test cases that more accurately reflect the performance of the fine-tuned LLMs. By integrating additional test cases from EvalPlus \cite{liu2024your}, EvalRepair-Java significantly increases the average number of test cases per problem from 7 to 583. This substantial expansion of test cases plays a vital role in mitigating the risk of patch overfitting caused by a limited number of test cases during the evaluation process.
\end{itemize}

\subsection{Evaluation Models}
\noindent\textbf{Selection Criteria.} To thoroughly evaluate the effectiveness and generalizability of federated fine-tuning, we select a diverse set of LLMs based on several crucial criteria:
\begin{itemize}
    \item[--] Firstly, it is imperative that the chosen LLMs are explicitly code-targeted, as they are pre-trained with code-related corpora to understand and generate programming languages, thereby enhancing the accuracy and effectiveness of code-related tasks.
    \item[--] Secondly, diversity in underlying model architectures and sizes is of vital importance, enabling exploration of the generalizability and practicality of existing LLMs. Due to the constraints imposed by computational resources, we select models with sizes ranging from 7B to 15B parameters.
    \item[--] Thirdly, we select LLM variants that are specifically fine-tuned for instruction following since such tailored versions of LLMs can capture complex prompts more accurately during fine-tuning.
    \item[--] Furthermore, we also take into consideration the prevalence of different LLMs, as widely recognized models benefit from extensive community support, facilitating better integration and troubleshooting. On the other hand, we can also obtain a more comprehensive understanding of popular LLMs. Therefore, we choose widely used LLMs of code based on the numbers of downloads from HuggingFace\footnote{\url{https://huggingface.co/models}}. We obtain the overall numbers of historical downloads for each model through the official API\footnote{\url{https://huggingface.co/api/models}} from HuggingFace.
    \item[--] Additionally, we also consider LLMs that have demonstrated superior performance in the current community to determine the final selection of the model. In this way, we can investigate whether existing state-of-the-art models can still maintain exceptional performance in our study. To that end, we refer to the leaderboard\footnote{\url{https://evalplus.github.io/leaderboard.html}} from EvalPlus \cite{liu2024your}, which provides a comprehensive evaluation of existing state-of-the-art LLMs of code to guide model selection for our study. Specifically, the EvalPlus leaderboard assesses the LLMs on its enhanced HumanEval \cite{chen2021evaluating} and MBPP \cite{austin2021program} benchmarks with expanded test cases that are 80 times and 35 times more than the original benchmarks, respectively. All the LLMs are ranked according to the average $Pass@1$ scores on the leaderboard.
\end{itemize}
     
Consequently, we take all the criteria into account and six different LLMs of code are selected in our study. The overview of the selected models is presented in Table \ref{llms}, including the model configurations, number of model downloads and average scores from the EvalPlus leaderboard.

\begin{itemize}
    \item[--] \textbf{CodeLlama.} CodeLlama \cite{roziere2023code} is a cutting-edge family of LLMs developed by Meta AI, specifically tailored for code generation and understanding tasks, and can generate code snippets based on surrounding context with infilling capabilities. Building on the Llama 2 \cite{touvron2023llama} architecture, CodeLlama includes several variants, such as the base model, Python-specific model, and instruction-following model. CodeLlama is designed for both research and commercial applications, making it a versatile tool for advancing code-related tasks in various domains. Due to resource constraints and the need to better understand the prompt, we employ CodeLlama-13B-Instruct and CodeLlama-7B-Instruct in our study to evaluate federated fine-tuning. 

    \item[--] \textbf{DeepSeekCoder.} DeepSeekCoder \cite{guo2024deepseek} is a series of open-source models specifically designed for coding tasks. The pre-training of DeepSeekCoder employs a project-level code corpus and a complementary fill-in-the-blank objective, enabling the model to handle code completion and infilling tasks at the project scale. DeepSeekCoder models excel in handling complex coding scenarios, outperforming existing closed-source models across various benchmarks. In this study, we employ DeepSeekCoder-7B-Instruct-V1.5 to leverage its advanced capabilities in code-related tasks.

    \item[--] \textbf{WizardCoder.} WizardCoder \cite{luo2023wizardcoder} is a state-of-the-art LLM of code that enhances code generation and comprehension through advanced instruction fine-tuning techniques. It utilizes the Evol-Instruct \cite{xu2023wizardlm} method that enables fine-tuning with more complex and diverse instructions for better code-related instruction following. We use WizardCoder-15B-V1.0 in this study to evaluate its applicability in federated learning.

    \item[--] \textbf{Mistral.} Mistral-7B \cite{jiang2023mistral} uses Mistral as the base model and has made progress in the attention mechanism, demonstrating the potential of auto-regressive models. Mistral-7B is able to outperform Llama 2 across all benchmarks. We employ the instruction-tuned version Mistral-7B-Instruct-v0.2, which provides a larger context window and no sliding-window attention compared with previous versions, to validate its effectiveness and consistency in the federated learning setting.

    \item[--] \textbf{CodeQwen.} CodeQWen \cite{bai2023qwen} is a specialized language model developed as part of the QWen series, designed to excel in code-related tasks. It builds upon the foundational QWen model and is a code-specific version of QWen, leveraging advanced training techniques such as reinforcement learning from human feedback (RLHF) to enhance its performance in coding tasks. We use CodeQwen1.5-7B-Chat, which is also an instruction-following version of CodeQWen, demonstrating robust code generation abilities and competitive performance across various code-related tasks including program repair.
\end{itemize}

For brevity, we refer to CodeLlama-13B-Instruct, CodeLlama-7B-Instruct, DeepSeekCoder-7B-Instruct-V1.5, WizardCoder-15B-V1.0, Mistral-7B-Instruct-v0.2, CodeQwen1.5-7B-Chat as CodeLlama-13B, CodeLlama-7B, DeepSeekCoder-7B, WizardCoder-15B, Mitral-7B and CodeQWen-7B in the subsequent sections of this paper.

\begin{table*}
\fontsize{7}{7}\selectfont
\centering
\caption{Overview of Selected LLMs of Code}
\begin{threeparttable}
\begin{tblr}{
  width = \linewidth,
  colspec = {Q[373]Q[183]Q[135]Q[142]Q[104]},
  cells = {c},
  hline{1-2,8} = {-}{},
}
Model                          & Base Model   & Model Size & \# Download & Average Score \\
CodeLlama-13B-Instruct         & Llama2       & 13B        & 601.0K      & 45.5\tnote{*}    \\
CodeLlama-7B-Instruct          & Llama2       & 7B         & 886.3K      & 41.1\tnote{*}    \\
DeepseekCoder-7B-Instruct-V1.5 & DeepSeek-LLM & 7B         & 115.5K      & 66.8     \\
WizardCoder-15B-V1.0           & CodeLlama    & 15B        & 155.9K      & 52.4    \\
Mistral-7B-Instruct-v0.2       & Mistral      & 7B         & 11.9M    & 36.5     \\
CodeQwen1.5-7B-Chat            & QWen         & 7B         & 115.8K      & 73.8     
\end{tblr}
\begin{tablenotes}
    \footnotesize
    \item[*] The instruction-tuned versions of CodeLlama-13B and CodeLlama-7B are not evaluated in the leaderboard, thus the scores of the base versions are provided as a reference.
\end{tablenotes}
\end{threeparttable}
\label{llms}
\end{table*}

\subsection{Evaluation Metrics}
\noindent$\boldsymbol{Top@k}$. To evaluate the performance of the fine-tuned LLMs, we employ the $Top@k$ metric, which assesses the model's ability to generate correct solutions within a limited number of attempts. The value of $k$ represents the number of samples generated for each problem. In the $Top@k$ evaluation, a problem is considered solved if at least one of the $k$ generated samples for that problem passes the corresponding test cases. The $Top@k$ score is then calculated as the proportion of successfully solved problems within the given k attempts. Previous research has shown that developers are less likely to utilize automated repair tools if they fail to generate a viable solution within a limited number of attempts \cite{kochhar2016practitioners, noller2022trust}. A study by Kochhar et al. \cite{kochhar2016practitioners} found that developers tend not to use automated repair tools if they cannot find a suitable fix within five attempts. Similarly, Noller et al. \cite{noller2022trust} reported that the maximum number of patches developers are willing to review is typically around 10. Taking these findings into consideration, we adopt the $Top@5$ and $Top@10$ metrics to evaluate the performance of the fine-tuned LLMs. Furthermore, we use $Top@1$, which is also a representative of the $Top@k$ family, in order to evaluate the LLMs in generating accurate solutions with minimal effort from the developers.

\noindent$\boldsymbol{Pass@k}$. In addition, we leverage an unbiased metric $Pass@k$ \cite{chen2021evaluating} to assess the repairing performance comprehensively. The unbiased estimator is calculated as follows:

\begin{equation}
    Pass@k=\mathbb{E}_{\text{problems}}\left[1 - \frac{\binom{n-c}{k}}{\binom{n}{k}}\right]
\end{equation}

\noindent where $n$ is the total number of generated samples by the LLM, $k\leq n$ is the number of attempts, and $c\leq n$ is the number of correct samples. We obtain $Pass@k$ for each problem and then the expectation is taken over all problems. This metric calculates the expected value of the proportion of problems for which at least one correct solution is generated within $k$ attempts from the $n$ generated samples. The $Pass@k$ metric provides a more nuanced view of the model's performance, as it takes into account the distribution of correct solutions among the generated samples, rather than simply considering the presence or absence of a correct solution within a fixed number of attempts. Together with $Top@k$, we use $Pass@5$ and $Pass@10$ to evaluate the capability of fine-tuned LLMs in bug fixing.

\subsection{Experimental Setup}\label{setup}
\noindent\textbf{Hardware \& Software Configuration.} We conduct the experiments on 4 Nvidia RTX 4090 GPUs, each with 24GB of graphical memory. The system runs with Intel(R) Xeon(R) Platinum 8352V CPU @ 2.10GHz and 480GB of RAM on Ubuntu 20.04. We employ the FederatedScope-LLM (FS-LLM) \cite{kuang2024federatedscope} package as the fundamental framework for federated learning. FS-LLM is an open-source federated learning package developed by Alibaba, enabling large-scale and efficient federated fine-tuning for LLMs. We implement several key components, including the construction of diverse heterogeneous code scenarios, the parameter-efficient fine-tuning technique, and the evaluation benchmark for the program repair task based on this framework to conduct a comprehensive study. All of our approaches are built based on Pytorch 1.10.0 and Cuda 11.3.

\noindent\textbf{Hyperparameters.} We refer to the settings for LLMs of code from the work of Jiang et al. \cite{jiang2023impact} as the guidance of hyperparameter configuration in our study. We use the paged AdamW \cite{adamw} optimizer, which is a variant of the AdamW optimizer designed to mitigate the occasionally out-of-memory issue on GPUs by allocating CPU RAM for the GPUs, with a learning rate of $1e^{-4}$ to update the weights. The batch size is set to one due to significant computational resources caused by large-scale models, and we set the max length of input tokens to 2048 to capture as much context as possible under the resource constraint. We fine-tune at most 30 epochs on each client and set the rounds of the global exchange of model parameters to 10. The fine-tuning process is regularized by an early stopping strategy that prevents excessive fine-tuning if the validation loss does not significantly decrease after 10 epochs. The rank and dropout for QLoRA are set to 32 and 0.05, respectively.

%% file: 5.experiments.tex
\section{Experiments \& Results}\label{experiment}
\subsection{RQ1: Effectiveness of Federated Fine-tuning}\label{ex-rq1}

\noindent\textbf{[Experiment Goal for RQ1]:} We aim to explore the effectiveness of federated learning in enhancing fine-tuning LLMs for program repair task. We compare the performance of federated learning with other fine-tuning scenarios and the original model, assessing its capability to improve model performance while protecting private data.

\noindent\textbf{[Experiment Design for RQ1]:} To validate the effectiveness of federated fine-tuning, we first compare federated learning with the original model, which is not fine-tuned to assess whether LLM fine-tuning can enhance program repair in the context of federated learning. We further investigate the efficacy of federated fine-tuning in improving performance while preserving data privacy by comparing it with other fine-tuning methods, i.e., local fine-tuning and centralized fine-tuning. In local fine-tuning, the models are fine-tuned within the local client, which has no access to datasets from other clients, and the model is updated locally without interacting with the central server. Therefore, we are able to validate to what extent federated learning can enhance performance with collaboration. We also compare it with centralized fine-tuning, which is regarded as the ideal situation for most traditional federated learning tasks since it pools all data together to fine-tune models regardless of data privacy or any other restrictions. It allows us to examine how close the performance of federated learning can be to centralized learning or whether it can outperform the centralized method. Comparing with the original model ensures that federated fine-tuning can indeed boost the LLM rather than introducing performance loss to the model. On the other hand, comparing with other fine-tuning methods provides insights into the potential of federated fine-tuning to enhance program repair through collaboration without compromising data privacy. In this RQ, we set the number of clients $K$ to 100 to simulate the federated learning system where 100 clients collaboratively fine-tune the LLMs (the detailed workflow can be referred to in Section \ref{ff}).
We leverage $Top@5$, $Top@10$, $Pass@1$, $Pass@5$ and $Pass@10$ to evaluate the fine-tuned LLMs in bug fixing.

\noindent\textbf{[Experimental Results for RQ1]:} Table \ref{rq1} presents the evaluation results of different fine-tuning methods on EvalRepair-Java. The best results for each metric are highlighted in bold and the second-best results are underlined. 

\begin{table*}[htbp]
\fontsize{6.5}{6.5}\selectfont
\centering
\caption{Experimental results of the original model (Original), local fine-tuning (Local), federated fine-tuning (FL) and centralized fine-tuning (Central) on EvalRepair-Java.}
\begin{tblr}{
  width = \linewidth,
  colspec = {Q[52]Q[52]Q[54]Q[54]Q[54]Q[56]Q[27]Q[52]Q[54]Q[54]Q[54]Q[56]Q[27]Q[52]Q[54]Q[54]Q[54]Q[56]},
  cells = {c},
  cell{1}{1} = {r=2}{},
  cell{1}{2} = {c=5}{0.27\linewidth},
  cell{1}{8} = {c=5}{0.27\linewidth},
  cell{1}{14} = {c=5}{0.27\linewidth},
  cell{7}{1} = {r=2}{},
  cell{7}{2} = {c=5}{0.27\linewidth},
  cell{7}{8} = {c=5}{0.27\linewidth},
  cell{7}{14} = {c=5}{0.27\linewidth},
  hline{1,7,13} = {-}{},
  hline{3,9} = {2-6,8-12,14-18}{},
  hline{6,12} = {1-6,8-12,14-18}{},
  colsep = {2pt}
}
Scenario  & CodeLlama-13B   &                &                &                &                &  & CodeLlama-7B   &                &                &                &                &  & DeepSeekCoder-7B &                &                &                &                \\
          & Top@5           & Top@10         & Pass@1         & Pass@5         & Pass@10        &  & Top@5          & Top@10         & Pass@1         & Pass@5         & Pass@10        &  & Top@5            & Top@10         & Pass@1         & Pass@5         & Pass@10        \\
Original  & 56.44           & 66.87          & 20.76          & 52.36          & 65.59          &  & \uline{53.37}  & 64.42          & \uline{25.72}  & 54.31          & 63.44          &  & 39.26            & 53.37          & 10.10          & 34.85          & 50.62          \\
Local & 53.99           & 55.83          & 32.37          & 53.85          & 59.78          &  & 46.63          & 53.37          & 29.68          & 49.70          & 54.64          &  & 20.25            & 31.90          & 6.89           & 23.04          & 35.19          \\
FL  & \uline{63.80}   & \textbf{76.69} & \textbf{39.31} & \textbf{68.81} & \textbf{76.88} &  & \textbf{62.58} & \textbf{71.17} & \textbf{31.48} & \textbf{62.35} & \textbf{71.21} &  & \uline{48.46}    & \uline{68.10}  & \uline{16.61}  & \uline{51.62}  & \uline{69.06}  \\
Central   & \textbf{66.87}  & \uline{72.39}  & \uline{33.55}  & \uline{66.04}  & \uline{73.64}  &  & 52.76          & \uline{65.03}  & 23.78          & \uline{54.57}  & \uline{65.71}  &  & \textbf{62.58}   & \textbf{82.21} & \textbf{26.33} & \textbf{66.45} & \textbf{79.84} \\
Scenario  & WizardCoder-15B &                &                &                &                &  & Mistral-7B     &                &                &                &                &  & CodeQWen-7B      &                &                &                &                \\
          & Top@5           & Top@10         & Pass@1         & Pass@5         & Pass@10        &  & Top@5          & Top@10         & Pass@1         & Pass@5         & Pass@10        &  & Top@5            & Top@10         & Pass@1         & Pass@5         & Pass@10        \\
Original  & 42.33           & \uline{60.12}  & 13.40          & 42.92          & 58.44          &  & \uline{44.17}  & 52.76          & \uline{19.82}  & \uline{43.52}  & \uline{52.60}  &  & 85.28            & 89.57          & 53.14          & 83.82          & 89.47          \\
Local & 41.72           & 47.85          & 21.99          & 43.14          & 49.78          &  & 26.38           & 36.81           & 12.76           & 25.68           & 35.52           &  & 71.17            & 79.14          & 50.97          & 73.82          & 79.22          \\
FL  & \textbf{56.44}  & \textbf{68.10} & \textbf{26.10} & \textbf{57.94} & \textbf{68.27} &  & \textbf{60.12} & \textbf{69.33} & \textbf{29.31} & \textbf{58.14} & \textbf{69.21} &  & \uline{87.12}    & \uline{90.18}  & \uline{56.11}  & \uline{85.03}  & \uline{90.30}  \\
Central   & \uline{50.31}   & \uline{60.12}  & \uline{16.80}  & \uline{48.10}  & \uline{60.93}  &  & 42.94          & \uline{53.37}  & 14.20          & 40.06          & 52.23          &  & \textbf{90.80}   & \textbf{92.64} & \textbf{59.37} & \textbf{88.65} & \textbf{92.98} 
\end{tblr}
\label{rq1}
\end{table*}

Firstly, the results show that federated fine-tuning outperforms each original model on the benchmark. Specifically, for the $Top@10$ metric, the improvement is most significant on the Mistral-7B model, reaching an increase of 16.57\%, while the smallest improvement is observed on the CodeLlama-7B model, with an increase of 6.75\%. For the $Pass@10$ metric, the greatest improvement is on the DeepSeekCoder-7B model, achieving a boost of 18.44\%, whereas the Codellama-7B model experiences the smallest increase of 7.77\%. These results indicate that federated fine-tuning can indeed enhance bug-fixing capabilities with LLMs. 

Secondly, the performance of federated fine-tuning surpasses local fine-tuning across all LLMs. We observe that local fine-tuning performs the worst among all fine-tuning methods, especially on DeepSeekCoder-7B where the reduction on $Top@10$ is up to 21.47\% compared to the original model. On the other hand, Mistral-7B shows the most significant decrease at 17.08\% in terms of $Pass@10$, highlighting the necessity of collaborative learning. 

The results also show that federated fine-tuning achieves the best scores in terms of both $Top@k$ and $Pass@k$ with CodeLlama-13B, CodeLlama-7B, WizardCoder-15B and Mistral-7B, which, to our surprise, even outperform those of centralized fine-tuning. Notably, the federated fine-tuning approach achieves a significant increase of 16.98\% for $Pass@10$ and 15.96\% for $Top@10$ on Mistral-7B compared with the centralized fine-tuning approach. The best performance is achieved by centralized fine-tuning for DeepSeekCoder-7B and CodeQWen-7B. However, federated fine-tuning achieves the second-best performance, approaching the central method, particularly for CodeQWen-7B. The differences between federated fine-tuning and centralized fine-tuning on CodeQWen-7B are only 2.46\% and 2.68\% for $Top@10$ and $Pass@10$, respectively.

Centralized learning is often regarded as the upper bound for traditional DL tasks in federated learning since it represents the ideal situation in real-world scenarios where all data are put together. Our results reveal that federated fine-tuning with LLMs can achieve comparable performance to centralized learning while protecting private data, whereas the central approach compromises data privacy. Besides, we also notice in a few previous research \cite{yang2024federated,jiang2023low, yue2023fedjudge} that the LLM-based federated fine-tuning approach has the potential to outperform centralized fine-tuning, revealing the unsuitability of simply combining data from different distributions for centralized learning.

We also speculate that it is because in traditional federated learning tasks, models like deep neural networks have to learn from scratch and centralized training has direct access to all data, thus performing better than federated learning in general cases. However, our results demonstrate that this advantage of the central method does not apply to fine-tuning models like LLMs since they have already been pre-trained with massive corpora.

\begin{tcolorbox}
\textbf{Finding 1:} Federated fine-tuning can effectively enhance program repair capabilities compared to other fine-tuning methods, even rivaling the centralized fine-tuning approach. Federated fine-tuning exhibits the potential for industries to collaborate and improve performance without compromising data privacy.
\end{tcolorbox}

We also note that while centralized fine-tuning generally performs well in most scenarios, a few cases still exist where the central approach even performs worse than the original models. For example, the $Top@5$ and $Pass@1$ of the centralized approach on CodeLlama-7B are 0.61\% and 1.94\% lower than the original model. In addition, centralized fine-tuning results in worse performance on Mistral-7B for all metrics except for $Top@10$ compared with the original model, with the $Pass@1$ metric notably decreasing by 5.62\%. Moreover, the local fine-tuning approach on all LLMs presents the worst performance among all baselines. Since the central approach utilizes the full dataset whereas the local approach fine-tunes the LLMs with only a portion of all data, we speculate that the size of the fine-tuning dataset could influence the performance of LLM fine-tuning, which will be further discussed in detail in Section \ref{disc}.

In summary, compared with the local and centralized fine-tuning approaches, federated fine-tuning enhances program repair by facilitating indirect client collaboration to utilize all datasets implicitly while protecting data privacy, thus combining the advantages and overcoming the disadvantages of both approaches. Furthermore, federated fine-tuning demonstrates promising and consistent performance across all LLMs, whereas the other two approaches exhibit performance degradation in some cases.

\begin{tcolorbox}
\textbf{Finding 2:} The local fine-tuning and centralized fine-tuning approaches demonstrate instability and inconsistency in fine-tuning LLMs for program repair. Despite that centralized fine-tuning achieves promising performance in most cases, there are still some LLMs fine-tuned by centralized learning that perform worse than the original models such as CodeLlama-7B and Mistral-7B. On the other hand, the local fine-tuning approach results in the worst performance compared to all baselines, which again highlights the advantage of federated fine-tuning and the necessity in collaboration. 
\end{tcolorbox}

\begin{figure}[htbp]
\centering
\includegraphics[width=0.7\textwidth]{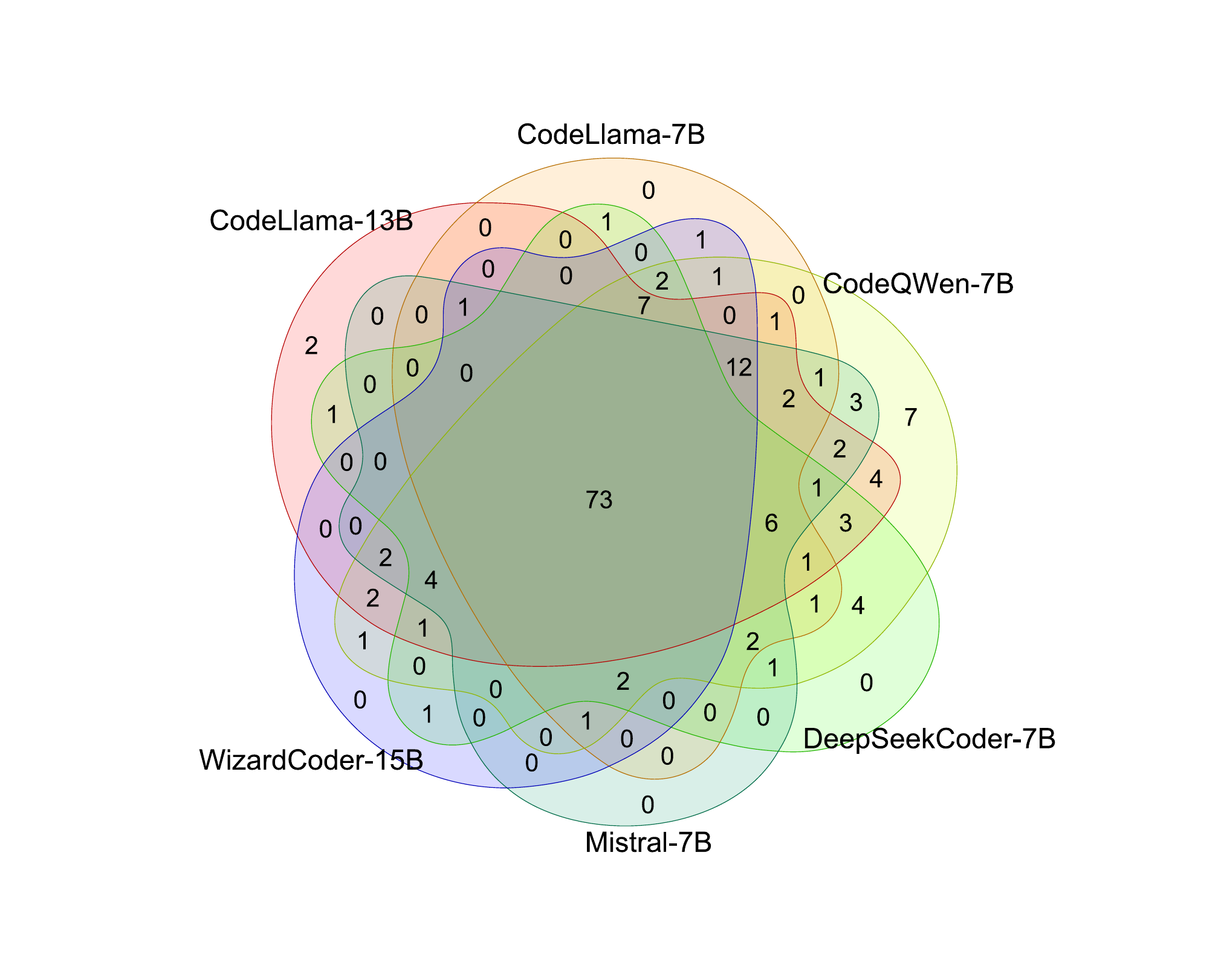}
\caption{Bug fix Venn diagram on EvalRepair-Java. \ye{impossible to read the numbers.}}
\label{venn}
\end{figure}

The Venn diagram of unique fixes generated by different LLMs with federated fine-tuning is presented in Figure \ref{venn}. Despite the 73 bugs repaired by all LLMs, CodeQWen-7B is able to fix 7 unique bugs, followed by CodeLlama-13B, which can fix 2 unique bugs. However, CodeLlama-13B exhibits a narrower scope in terms of the bugs it can fix, with a relatively smaller overlap with other LLMs compared with CodeQWen-7B. This emphasizes the complementary strengths of CodeQWen-7B since it presents larger overlaps of fixes with and without CodeLlama-13B. For example, CodeQWen-7B is able to fix 3 bugs with Mistral-7B, 4 bugs with DeepSeekCoder-7B, and 4 bugs with CodeLlama-13B, which demonstrates that CodeQWen-7B can be complementary to other LLMs of code, presenting the potential of CodeQWen-7B for further use.

\begin{tcolorbox}
\textbf{Finding 3:} CodeQWen-7B and CodeLlama-13B are able to fix 7 and 2 more unique bugs than others, respectively. It is noteworthy that CodeQWen-7B demonstrates its complementary strength in relation to CodeLlama-13B. Not only does CodeQWen-7B exhibit a larger overlap with CodeLlama-13B, but it also extends its bug-fixing capabilities beyond the scope covered by CodeLlama-13B, indicating that CodeQWen-7B complements CodeLlama-13B and other LLMs by addressing a broader range of bugs.
\end{tcolorbox}

\subsection{RQ2: Impact of data heterogeneity}

\noindent\textbf{[Experiment Goal for RQ2]:} We aim to further investigate the impact of various data distributions on LLM-based federated fine-tuning for program repair. We construct heterogeneous code to explore whether fine-tuning LLMs under diverse Non-IID scenarios is able to enhance the performance and whether it incurs negative impacts on the performance compared to IID scenario.

\noindent\textbf{[Experiment Design for RQ2]:} In order to comprehensively analyse the impact of Non-IID data on fine-tuning LLMs for program repair, we construct different degrees of Non-IID scenarios along with the IID scenario to evaluate federated fine-tuning. Most previous research in federated learning and applying federated learning to SE mainly focuses on typical supervised tasks such as classification, regression, etc, which enable them to construct Non-IID data conforming to label skew based on the labeled datasets. However, few of them pay attention to generative tasks such as program repair and other code-related tasks that involve feature skew, which is also a critical challenge that may occur in real-world applications because of different software development practices across companies. Existing research constructs feature skew by collecting datasets from different data sources, domains, or individuals that inherently exhibit Non-IID properties \cite{li2021fedbn,li2024fedcir,tan2023fedsea,chen2023fraug}. The code repositories from different companies can differ to varying degrees because they follow different coding standards, developers have different levels of programming skills, or they need to satisfy different business requirements. Therefore, we leverage three types of code features in our study to construct feature-skewed Non-IID data to study its impact in depth. 

\textbf{Coding Style.} Firstly, considering program readability and maintainability, companies generally would enforce specific programming guidelines and requirements to restrict coding style for easier code modification or code refactoring in the future \cite{ting2023codestylist,wiese2019replicating}. Individuals can write the same program in various ways because of different programming habits. For example, the usage of Camel case, Pascal case, or underscores as identifier naming method, the preference to use \textit{if} or \textit{switch} for conditional structures, etc. Coding style is primarily composed of four different attributes, namely layout features, lexical features, syntactic features, and semantic features \cite{caliskan2015anonymizing,li2022ropgen}. A brief overview of the four types of coding style attributes is presented in Table \ref{csa}. Previous works extract coding styles in source code authorship attribution to identify programmers, students, or other individuals \cite{azcona2019user2code2vec,caliskan2015anonymizing,kovalenko2020building}. We use coding style as the first type of code feature to construct feature-skewed Non-IID data. In order to obtain the coding style attributes from the dataset, we follow the work in \cite{li2022ropgen} to extract the attributes from our dataset, which results in 23 specific attributes across lexical, syntactic, and semantic features.

\begin{table*}[htbp]
\fontsize{7}{7}\selectfont
\centering
\caption{A brief overview of coding style attributes.}
\begin{tblr}{
  width = \linewidth,
  colspec = {Q[119]Q[388]Q[433]},
  cells = {c},
  hline{1-2,6} = {-}{},
}
Attribute Type & Description                                          & Example                                                    \\
Layout         & The organization of code                             & Usage of indention, empty lines, etc.                      \\
Lexical        & The tokens used in the code                          & Usage of naming method, variable, global declaration, etc. \\
Syntactic      & Tree structure and syntactic constructs of a program & Location of variables, variable assignment, etc.           \\
Semantic       & Control flows and data flows of a program            & Usage of loop structures, conditional structures, etc.     
\end{tblr}
\label{csa}
\end{table*}

\textbf{Code Complexity.} Secondly, code complexity is one of the major aspects that affects the internal quality of software during software development and complex code can increase the cognitive burden for code comprehending \cite{antinyan2017evaluating,peitek2021program}. However, developers from different enterprises can vary in their experience, problem-solving approaches, and programming skills, which generally end up with different complexity of code. Therefore, we consider code complexity as the second code feature to construct feature-skewed Non-IID data. According to Antinyan et al. \cite{antinyan2017evaluating}, there are three direct sources of code complexity: (1) The elements and connections existing in the code. (2) Representational clarity of the code. (3) Intensity of evolution. The third source indicates that the frequency and magnitude of changes in the code can directly reflect the complexity level of the code since it influences the time cost of code maintenance. In view of this, to assess the code complexity of our bug-fix code pairs, we leverage the number of modified code hunks of each pair by comparing the buggy code with its fixed version to indicate the complexity levels of our dataset.

\textbf{Code Embedding.} Thirdly, we extract code embeddings to capture the underlying semantic and syntactic information of code. Due to the fact that different companies engage in diverse business activities, they inevitably have varied software development needs. This diversity leads to underlying differences in the information presented within their codebases. These distinct software requirements also reveal differing intentions within the code, reflecting the unique objectives of each company. Therefore, we extract the code embeddings from our dataset as the third code feature to construct feature-skewed Non-IID data since code embeddings provide an enhanced understanding of the semantic and syntactic meanings of code tokens as the pre-trained models are developed using extensive external datasets \cite{ding2022can}. Furthermore, we utilize CodeBERT \cite{feng2020codebert} as the pre-trained model for code embedding extraction. CodeBERT is capable of capturing the semantic relationship between natural language (NL) and programming language (PL) and can effectively support NL-PL understanding for generative tasks. Given that our dataset provides natural language descriptions as metadata for each code, we use the NL-PL pair as the input for CodeBERT and obtain the contextual vector representation as the output.

\textbf{Feature Clustering.} Lastly, we aim to construct feature-skewed Non-IID scenarios based on the three code features to further explore the impacts of data heterogeneity. Existing studies construct feature-skewed Non-IID data mainly through feature probability distribution and code clustering based on embeddings \cite{nguyen2022fedprob,lin2021fednlp}. Similar to the idea of label-skewed scenarios where data samples are allocated to clients in a Non-IID manner according to the label distribution of the dataset, clustering unlabeled data into groups provides a form of implicit labels that can guide the Non-IID data allocation. Previous studies cluster the code by extracting code description features, contextual syntactic features, code change features, or using a code embedding model to encode the code \cite{ye2021automated,liang2023needle,kreutzer2016automatic}. Considering real-world scenarios as described previously, we perform code clustering based on the coding style and code embedding features except for code complexity, of which we use the complexity level as the label. We adopt Bisecting-K-means \cite{michael2000comparison}, which is a hierarchical clustering algorithm in order to determine a reasonable number of clusters. To that end, we analyse the sum of squared error (SSE) to determine the appropriate number of clusters. The SSE is defined as follows:

\begin{equation}
    \label{e-sse}
    SSE=\sum_{k=1}^{K}\sum_{i\in C_k}(x_i-\mu_k)^2
\end{equation}

\noindent where $K$ is the number of clusters and $C_k$ is the set of elements in cluster $k$. $x_i$ represents the $i$-th element in cluster $k$ and $\mu_k$ is the centroid of cluster $k$. SSE measures the variance within the clusters, thus a smaller SSE indicates better clustering. 

\begin{figure*}[htbp]
    \centering
    \begin{subfigure}{0.49\textwidth}
        \centering        
        \includegraphics[width=\textwidth]{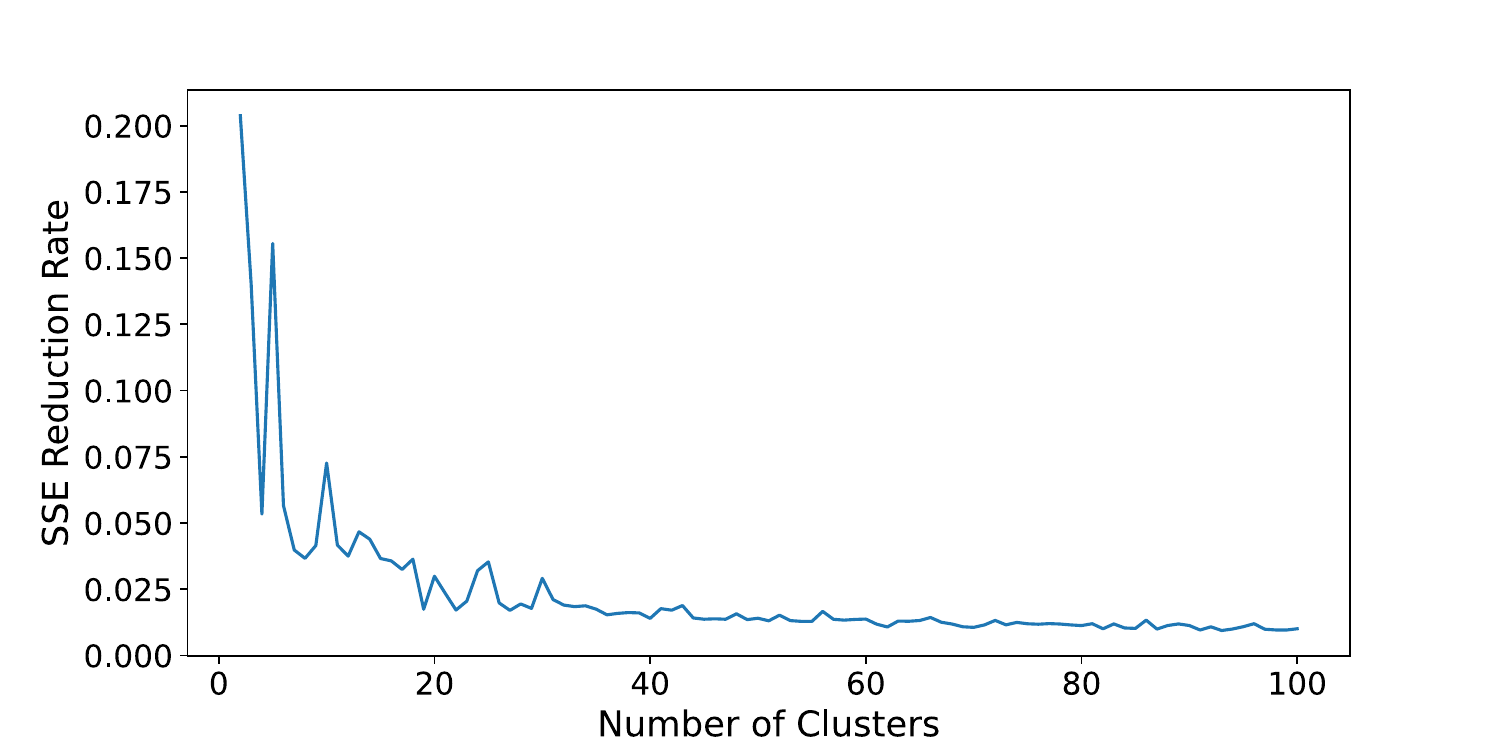}
        \caption{Clustering of coding style.}
        \label{cs-sse}
    \end{subfigure}
    \begin{subfigure}{0.49\textwidth}
        \centering
        \includegraphics[width=\textwidth]{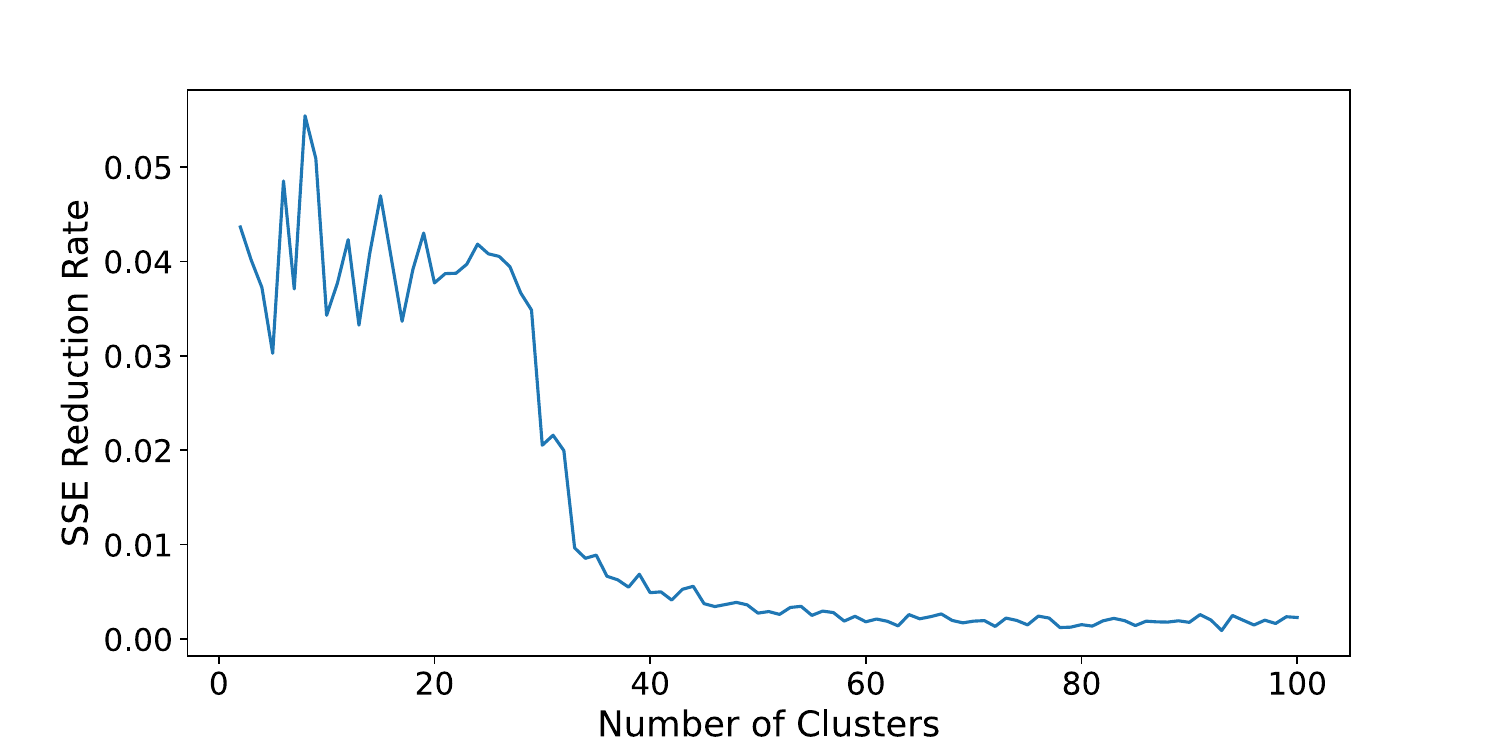}
        \caption{Clustering of code embedding.}
        \label{ce-sse}
    \end{subfigure}
\caption{SSE drop rate for different numbers of clusters.}
\label{sse}
\end{figure*}

\textbf{Optimizing Cluster Sizes.} Figure \ref{cs-sse} and Figure \ref{ce-sse} illustrate the reduction rate of SSE according to different numbers of clusters in terms of coding style and code embedding features. We observe that the SSE no longer declines significantly after the number of 20 in terms of the coding style feature. Due to resource constraints caused by clustering on high-dimensional data with a large number of clusters, we set the number of clusters to 40 for coding style according to the SSE drop rates. As demonstrated in Figure \ref{ce-sse}, the SSE drop rate decreases rapidly from approximately 25 to 40 clusters with a notably low rate of around 35 clusters, which is the number of distinct problems in TutorLLMCode. Since CodeBERT extracts the contextual information from both natural language description and code in the NL-PL pairs, underlying semantic information in the code can also be captured beside each unique problem. Similar purposes can exist in the code snippets that address different problems when they have common sub-problems, such as sorting and depth-first search, which results in lower SSE drop rates with more than 35 clusters. Taking into account the resource limitations, we set the number of clusters to 45 for the feature of code embedding. 

\begin{figure*}[htbp]
    \centering
    \begin{subfigure}{0.49\textwidth}
        \centering        
        \includegraphics[width=\textwidth]{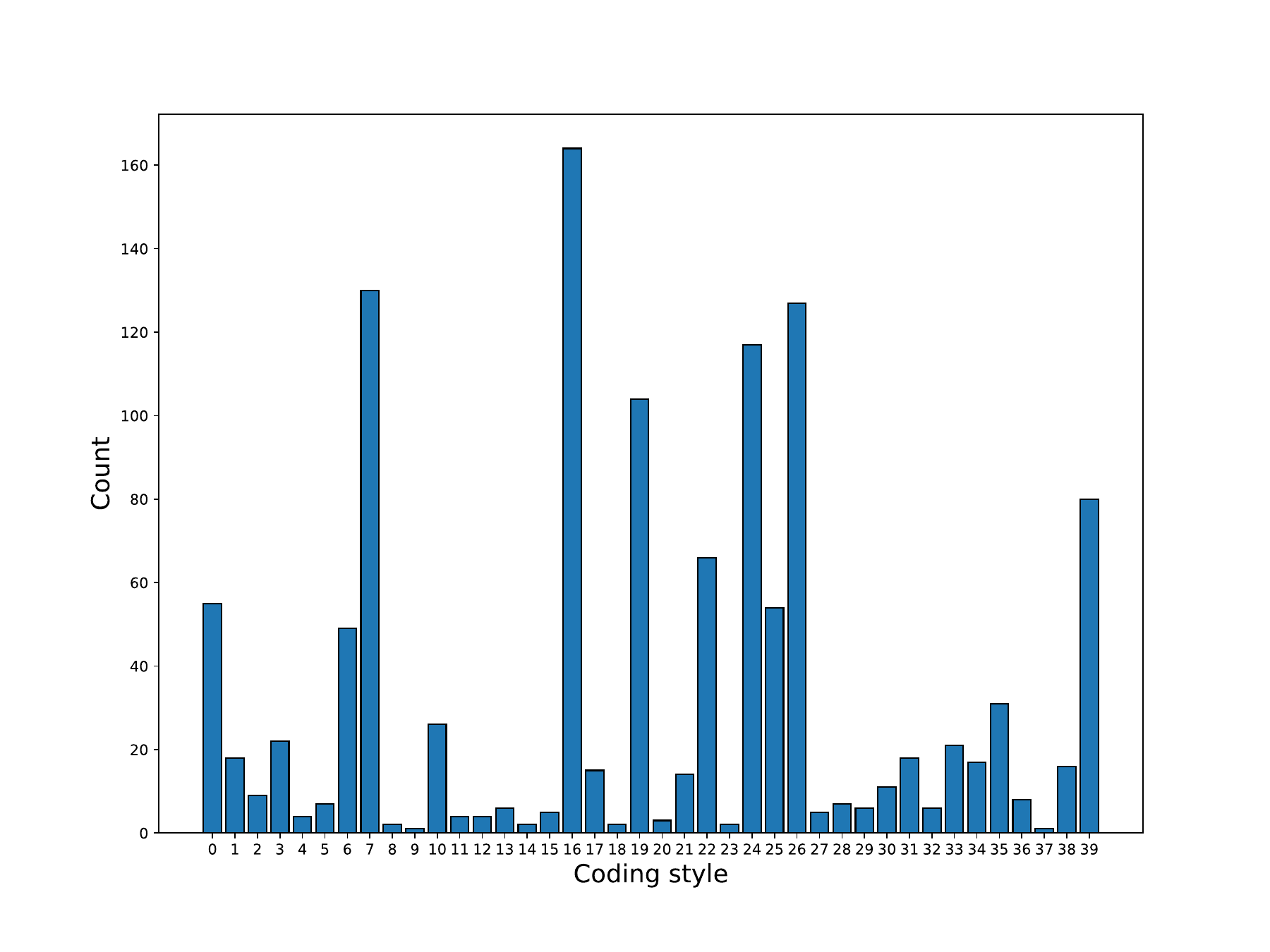}
        \caption{Distribution of coding styles.}
        \label{cs-hist}
    \end{subfigure}
    \begin{subfigure}{0.49\textwidth}
        \centering
        \includegraphics[width=\textwidth]{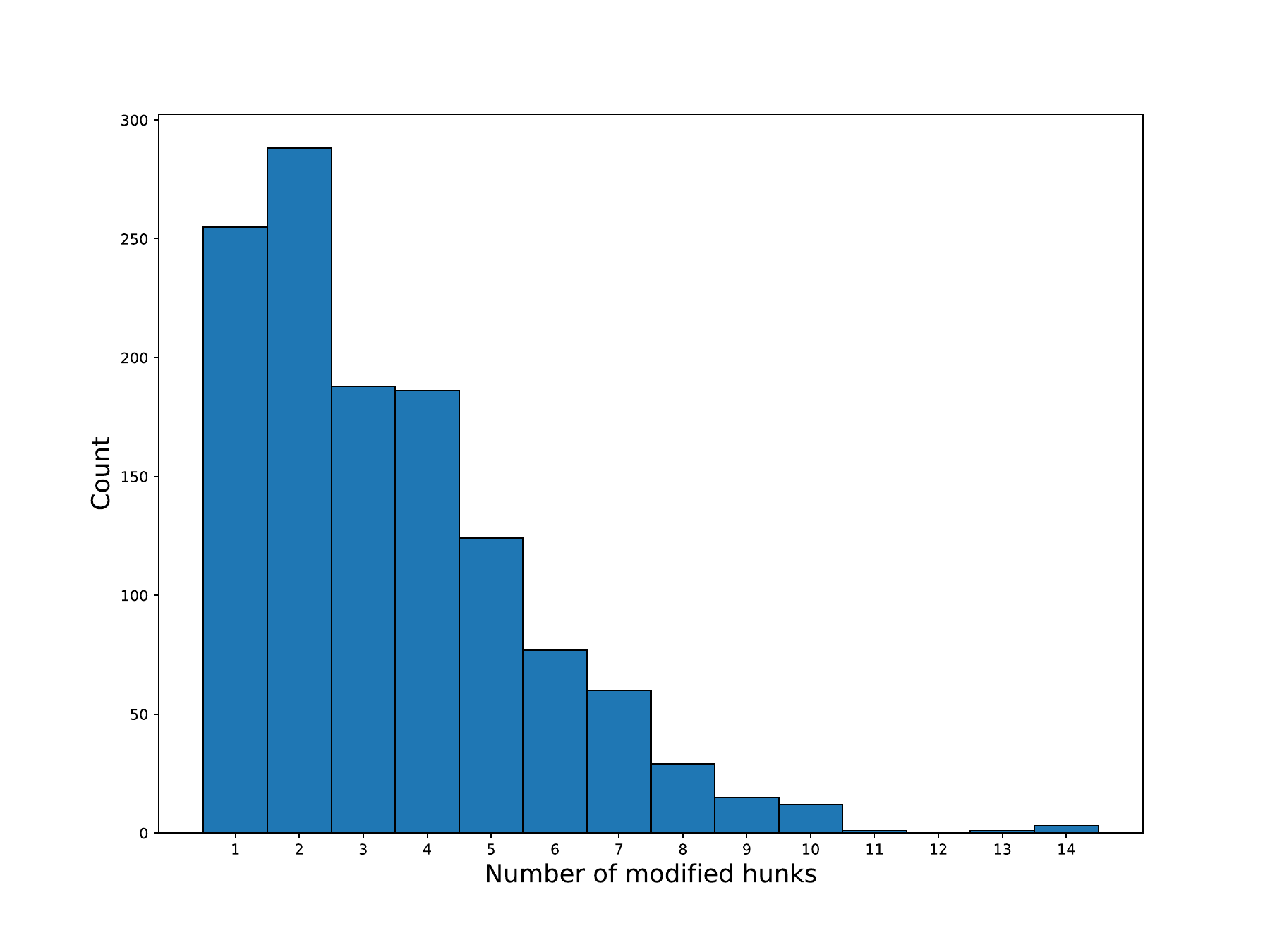}
        \caption{Distribution of code complexity.}
        \label{cc-hist}
    \end{subfigure}
\caption{Distribution of coding styles and code complexity.}
\label{hist}
\end{figure*}

Figure \ref{hist} presents the distribution of the 40 clustered coding styles and the distribution of different code complexity levels represented by the number of modified hunks. As depicted in Figure \ref{cc-hist}, there are 13 levels of code complexity since there is no code with 12 modified hunks. 

\textbf{Construction of Heterogeneous Code.} We allocate the datasets to each client in a Non-IID manner through the Dirichlet distribution \cite{huang2005maximum}, which is widely applied in previous studies \cite{li2022federated,kim2022multi,wu2023personalized}. Given a set of $N$ clients and $K$ clusters, the cluster assignments are assumed to follow a categorical distribution. Specifically, each client $n \in \{1, ..., N\}$ is assigned to a cluster according to a probability vector $\boldsymbol{q}_n$, where $\boldsymbol{q}_{n,k}>0$ denotes the probability that client $n$ belongs to cluster $k \in \{1, ..., K\}$ and $\lvert\lvert\boldsymbol{q}_n\rvert\rvert_1=1$. The probability vector $\boldsymbol{q}_n$ is drawn independently from a Dirichlet distribution, denoted as $\boldsymbol{q}_n\sim Dir(\alpha\cdot\boldsymbol{p})$ where $\alpha>0$ is a concentration parameter controlling the degree of heterogeneity in the cluster assignments and $\boldsymbol{p}$ represents a prior uniform distribution over the $K$ clusters. As $\alpha\rightarrow \infty$, the distributions of clusters assigned to each client are almost uniform, which also indicates the IID scenario. On the contrary, as $\alpha\rightarrow 0$, the distribution tends towards an extreme case, concentrating the probability mass on a single cluster for each client. Therefore, we construct different degrees of Non-IID scenarios by controlling $\alpha$. Specifically, we set $\alpha$ to 0.1 and 0.01 for mild and medium degrees of heterogeneity, respectively. In addition, we construct an extreme Non-IID scenario where $\alpha\rightarrow 0$ and each client only holds data of a single cluster. Therefore, we set the numbers of clients to 40 and 45 for coding style and code embedding, respectively, to fit the extreme Non-IID scenario according to their numbers of clusters. We set the number of clients to 10 in terms of code complexity because there are only a few or no code samples with numbers of modified hunks greater than 10, as presented in Figure \ref{cc-hist}. Therefore, we combine code samples with a number of modified hunks not less than 10 into a single cluster for better construction of the extreme Non-IID scenario. 

\begin{figure*}[htbp]
    \centering
    \begin{subfigure}{0.49\textwidth}
        \centering        
        \includegraphics[width=\textwidth]{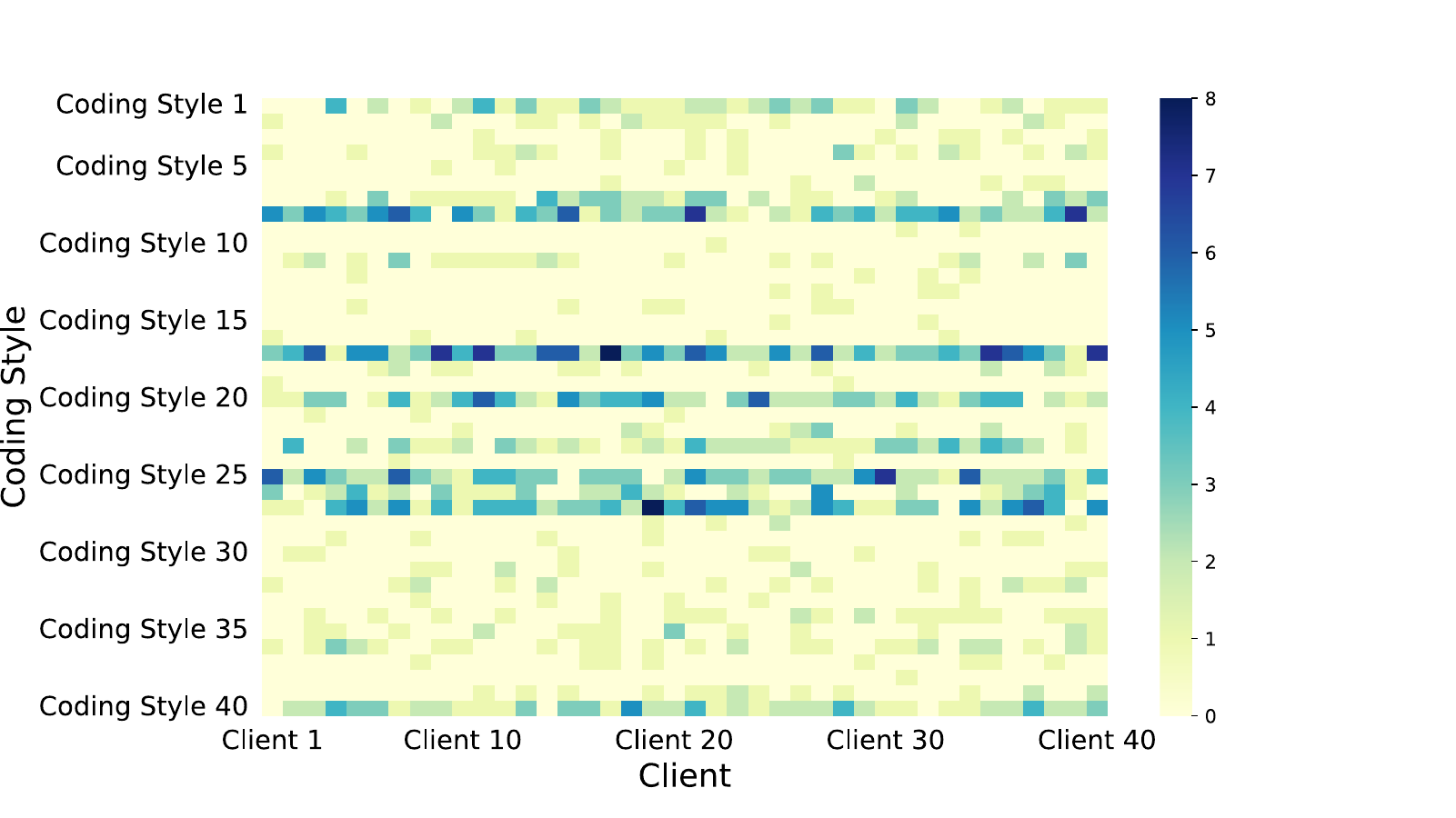}
        \caption{IID ($\alpha \rightarrow \infty$).}
        \label{iid}
    \end{subfigure}
    \begin{subfigure}{0.49\textwidth}
        \centering
        \includegraphics[width=\textwidth]{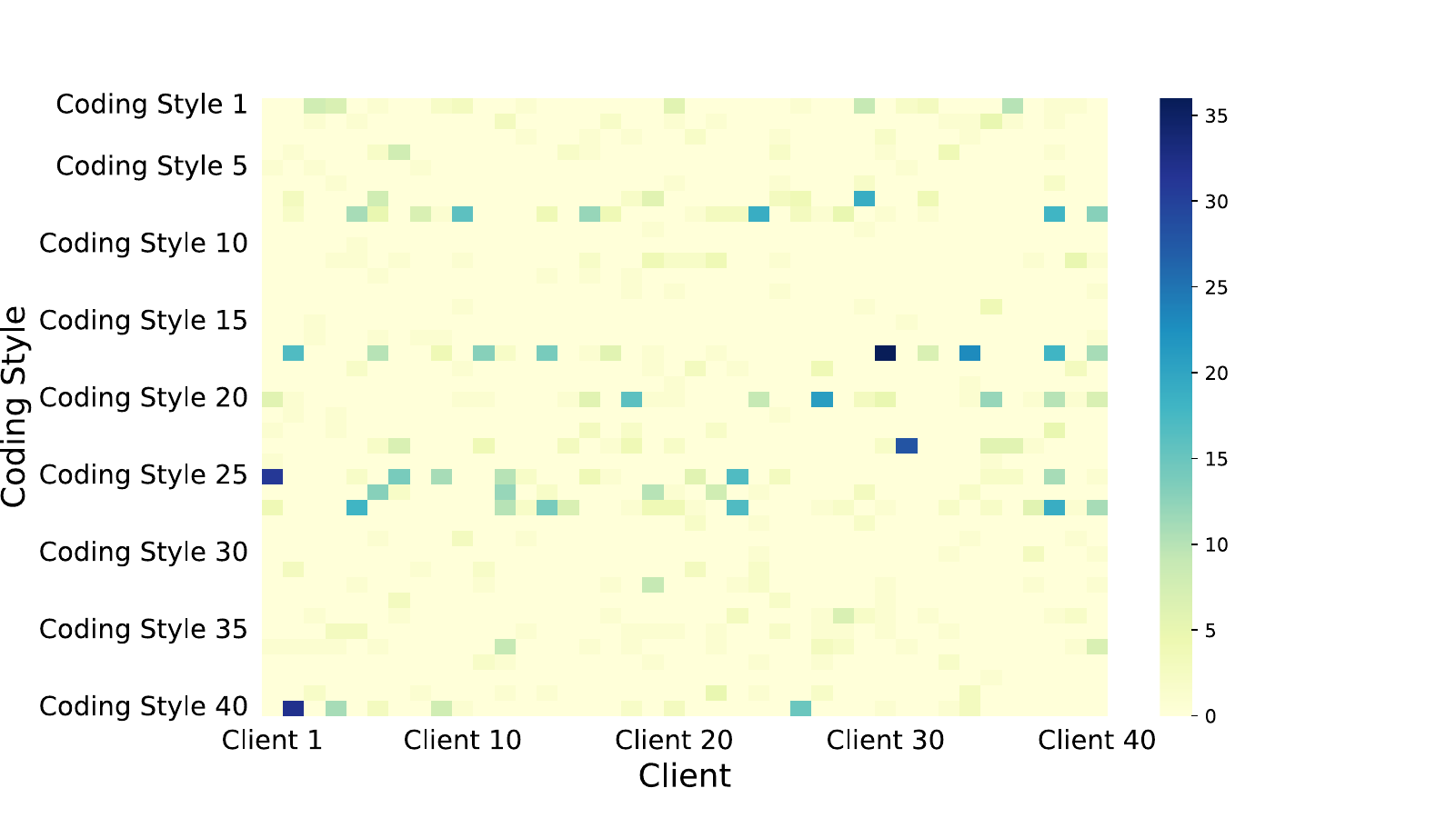}
        \caption{Mild Non-IID ($\alpha = 0.1$).}
        \label{mild}
    \end{subfigure}
    \begin{subfigure}{0.49\textwidth}
        \centering        
        \includegraphics[width=\textwidth]{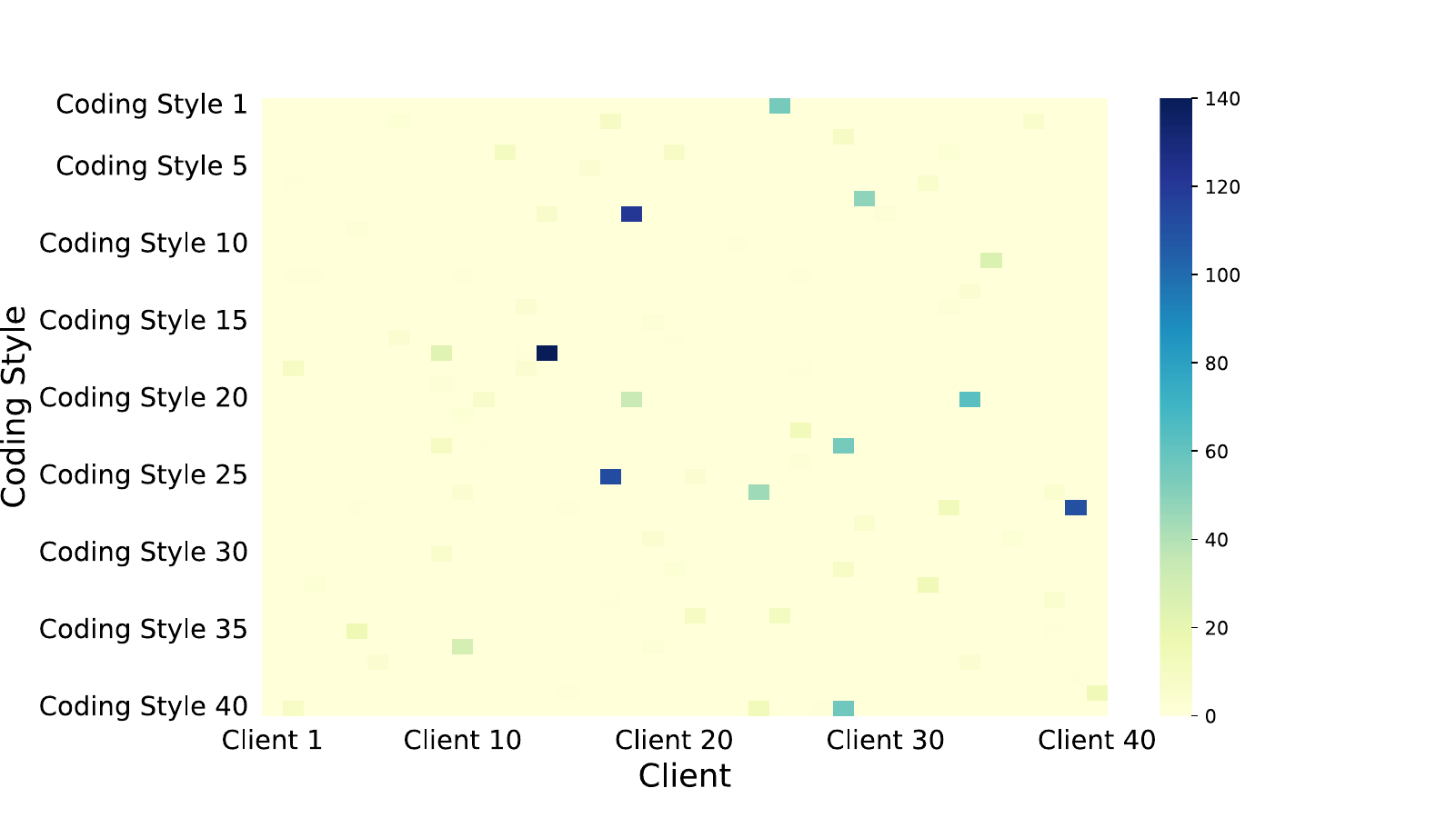}
        \caption{Medium Non-IID ($\alpha = 0.01$).}
        \label{medium}
    \end{subfigure}
    \begin{subfigure}{0.49\textwidth}
        \centering        
        \includegraphics[width=\textwidth]{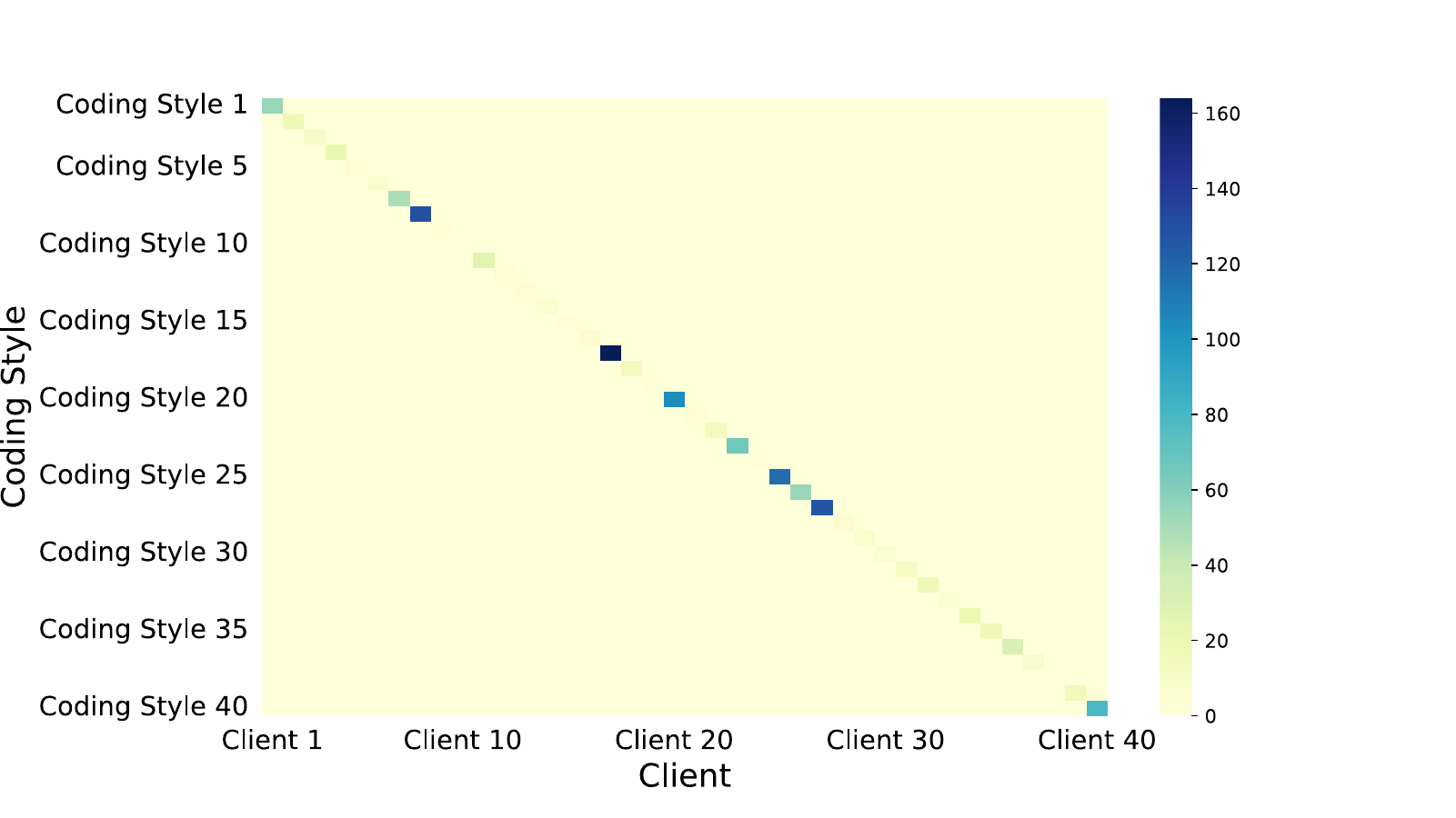}
        \caption{Extreme Non-IID ($\alpha \rightarrow 0$).}
        \label{ex}
    \end{subfigure}
    \caption{Different degrees of Data Heterogeneity.}
    \label{dist}
\end{figure*}

Figure \ref{dist} presents as an example the four data distributions employed in our study based on the feature of coding style. As we can observe from Figure \ref{iid}, the data distributions across clients tend to be identical, whereas the distributions in Figure \ref{mild}, Figure \ref{medium}, and Figure \ref{ex} become more concentrated as the degree of heterogeneity increases. In particular, each client possesses only a single cluster of data in the extreme Non-IID scenario, as shown in Figure \ref{ex}. We compare the four scenarios of data distribution with the original model and centralized fine-tuning to explore the impact of data heterogeneity. 

\begin{figure*}[htbp]
    \centering
    \includegraphics[width=1\textwidth, trim=1.01cm 16.3cm 1.01cm 0cm, clip]{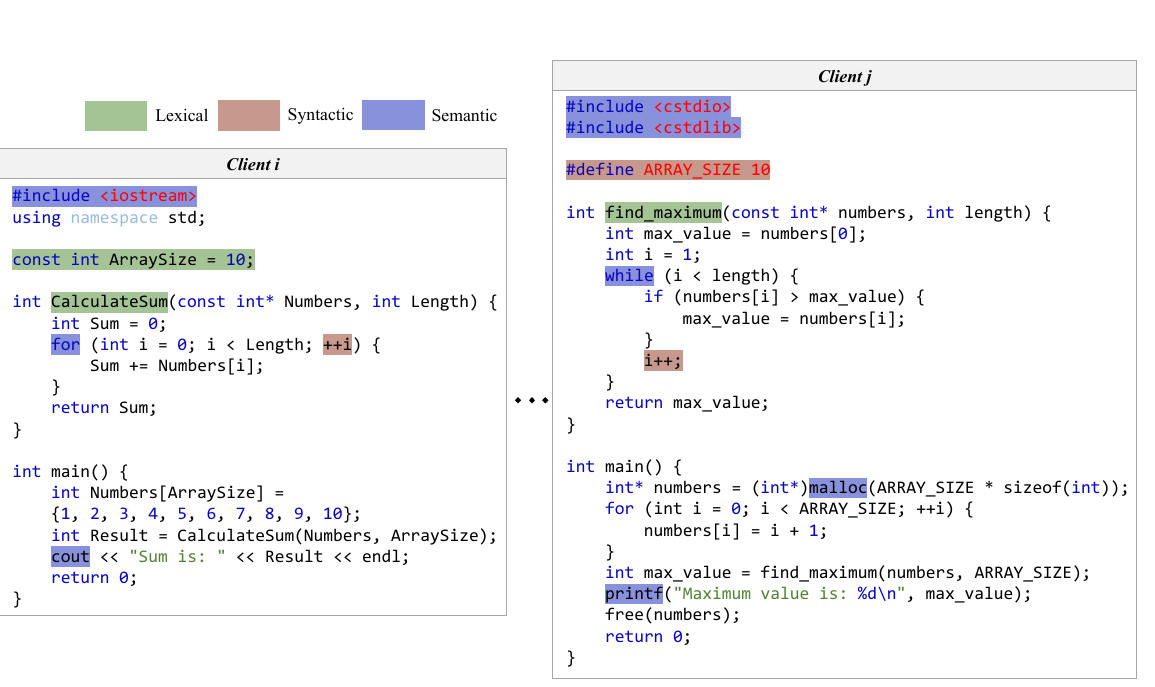}
    \caption{An example of heterogeneous coding style.}
    \label{cs-eg}
\end{figure*}

\begin{figure*}[htbp]
    \centering
    \includegraphics[width=1\textwidth, trim=1.01cm 19cm 1.01cm 0cm, clip]{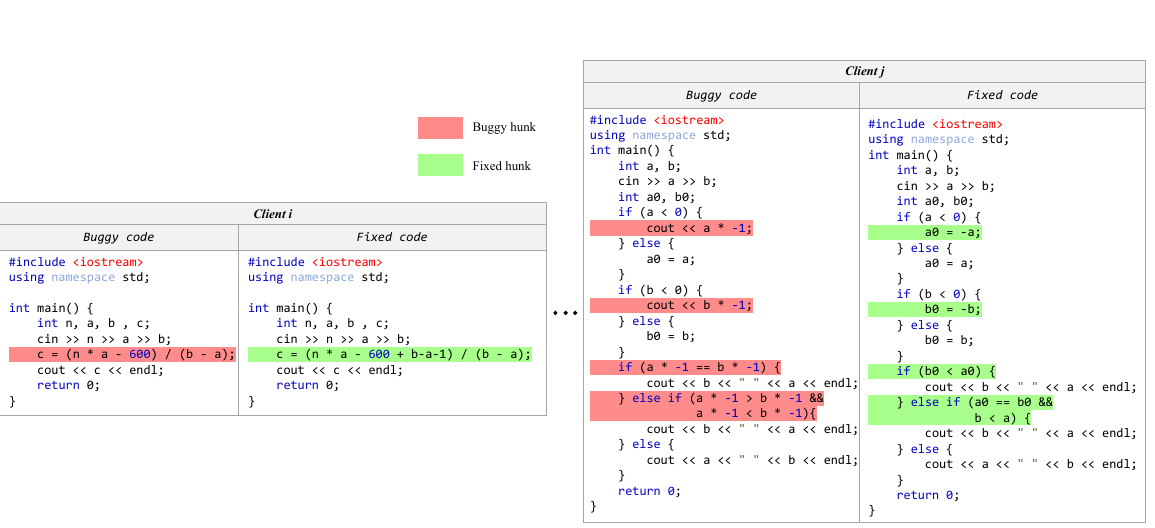}
    \caption{An example of heterogeneous code complexity.}
    \label{cc-eg}
\end{figure*}

\begin{figure*}[htbp]
    \centering
    \includegraphics[width=1\textwidth, trim=1.01cm 16.3cm 1.01cm 0cm, clip]{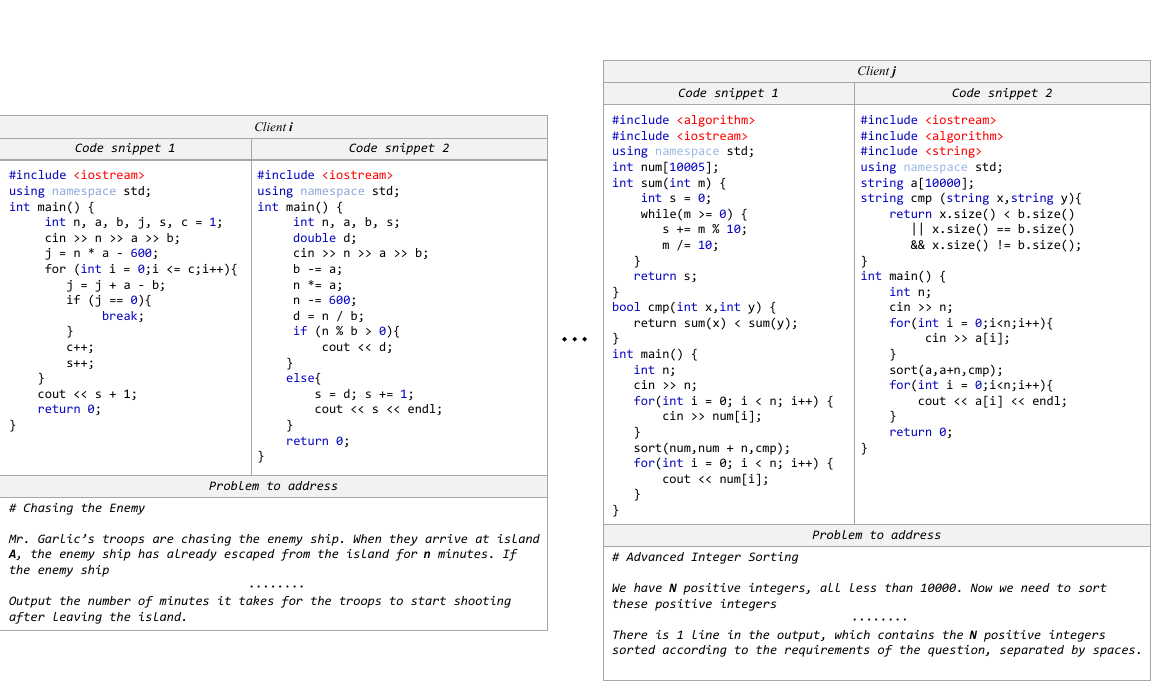}
    \caption{An example of heterogeneous code embedding.}
    \label{ce-eg}
\end{figure*}

\textbf{Examples of Heterogeneous Code.} The examples of heterogeneous coding style, code complexity, and code embedding distributed in different clients are presented in Figure \ref{cs-eg}, Figure \ref{cc-eg}, and Figure \ref{ce-eg}. We interpret the three examples as follows: (1) As depicted in Figure \ref{cs-eg}, the coding style of the two code snippets on client $i$ and client $j$ where $i\neq j$ differs across lexical, syntactic, and semantic attributes. The code snippet on client $i$ uses Pascal case as the naming method, whereas the method name is separated by an underscore in client $j$ in terms of lexical differences. The two clients also differ in the locations of increment operators, which is one of the syntactic differences. For semantic differences, client $i$ and client $j$ import different header files and use different loop structures. In addition, client $i$ uses static array allocation while client $j$ uses dynamic memory allocation. Assuming that each client in the federated learning system represents a different company, developers in real-world settings are likely to have varying coding styles as aforementioned. Thus, we evaluate heterogeneous coding styles to study their impacts on our work. (2) Figure \ref{cc-eg} demonstrates the heterogeneity in code complexity where client $i$ only needs to fix a single-hunk bug and client $j$ needs to fix four hunks of bugs. As a major factor affecting software maintenance, code complexity differs across code repositories from various industries, and the impact on LLMs learning with heterogeneous code complexity is to be evaluated in our study. (3) Since the contextual information is extracted by CodeBERT from the NL-PL pairs, the heterogeneity in code embedding, as illustrated in Figure \ref{ce-eg}, is reflected in different code snippets within a client addressing the same problem, whereas different clients target different goals. Real-world industries focus on addressing different problems due to a diversity of business requirements and objectives. Therefore, the impact of heterogeneous code embedding will also be explored in this study.

\textbf{Significance Test.} Furthermore, we employ Wilcoxon signed-rank test \cite{wilcoxon1992individual} to evaluate the statistical significance of the performance differences between various scenarios, in which the null hypothesis is that there is no significant difference between two scenarios. This non-parametric test is widely used by previous studies \cite{yang2024federated,yang2024large,zhang2023fedslice} and is particularly suitable for non-normal distributions. 

\textbf{Effect Size.} To further validate the significance of the observed differences and assess the magnitude of the effect, we adopt Cliff's delta \cite{cliff1993dominance}, a non-parametric effect size measure. Cliff's delta has been widely used in recent studies \cite{yang2024federated,schafer2023empirical,ahmed2022multilingual} to quantify the size of differences. The value of Cliff's delta is confined to the range $[-1,1]$, with 0 indicating no difference and values closer to -1 or 1 indicating a larger effect size. We follow the study of Yang et al. \cite{yang2024federated} to interpret the absolute value of Cliff's delta in different ranges as follows: (1) Negligible effect: $[0, 0.11)$. (2) Small effect: $[0.11,0.28)$. (3) Medium effect: $[0.28,0.43)$. (4) Large effect: $[0.43,1]$. 

Note that the numbers of clients set in this RQ are different from RQ1, which can produce different results for the same baselines.



\noindent\textbf{[Experimental Results for RQ2]:} The evaluation results of different data distributions based on different code features are presented in Table \ref{cs}, Table \ref{cc}, and Table \ref{ce}. Mild, Medium, and Extreme represent three different degrees of Non-IID scenarios as previously designed. Note that the best and second-best performances are highlighted only within the IID, Mild, Medium, and Extreme scenarios since we aim to investigate the impact of different data heterogeneity. We find that fine-tuning LLMs with heterogeneous code can still enhance the repairing capability and achieve promising results.

\begin{table*}[htbp]
\fontsize{6.5}{6.5}\selectfont
\centering
\caption{Experimental results of different coding style distributions on EvalRepair-Java.}
\begin{tblr}{
  width = \linewidth,
  colspec = {Q[54]Q[52]Q[54]Q[54]Q[54]Q[56]Q[27]Q[52]Q[54]Q[54]Q[54]Q[56]Q[27]Q[52]Q[54]Q[54]Q[54]Q[56]},
  cells = {c},
  cell{1}{1} = {r=2}{},
  cell{1}{2} = {c=5}{0.27\linewidth},
  cell{1}{8} = {c=5}{0.27\linewidth},
  cell{1}{14} = {c=5}{0.27\linewidth},
  cell{9}{1} = {r=2}{},
  cell{9}{2} = {c=5}{0.27\linewidth},
  cell{9}{8} = {c=5}{0.27\linewidth},
  cell{9}{14} = {c=5}{0.27\linewidth},
  hline{1,9,17} = {-}{},
  hline{3,11} = {2-6,8-12,14-18}{},
  hline{4,8,12,16} = {1-6,8-12,14-18}{},
  colsep = {2pt}
}
Scenario & CodeLlama-13B   &                &                &                &                &  & CodeLlama-7B   &                &                &                &                &  & DeepSeekCoder-7B &                &                &                &                \\
         & Top@5           & Top@10         & Pass@1         & Pass@5         & Pass@10        &  & Top@5          & Top@10         & Pass@1         & Pass@5         & Pass@10        &  & Top@5            & Top@10         & Pass@1         & Pass@5         & Pass@10        \\
Original & 56.44           & 66.87          & 20.76          & 52.36          & 65.59          &  & 53.37          & 64.42          & 25.72          & 54.31          & 63.44          &  & 39.26            & 53.37          & 10.10          & 34.85          & 50.62          \\
IID      & \uline{70.55}   & \uline{80.37}  & 36.01          & \uline{70.78}  & 79.26          &  & 57.06          & 64.42          & 27.51          & 58.76          & 67.43          &  & 43.56            & 60.12          & 11.84          & 42.10          & 61.72          \\
Mild     & \uline{70.55}   & 78.53          & 37.94          & 69.24          & 76.80          &  & \textbf{61.96} & \textbf{74.85} & 29.78          & \textbf{62.76} & \textbf{73.30} &  & \uline{52.76}    & 68.10          & 15.95          & 50.04          & 67.46          \\
Medium   & 69.33           & 74.23          & \uline{38.18}  & 70.48          & \uline{80.72}  &  & \uline{60.74}  & \uline{69.33}  & \uline{37.05}  & \uline{62.30}  & \uline{69.14}  &  & \textbf{58.90}   & \textbf{75.46} & \textbf{22.22} & \textbf{60.78} & \textbf{76.24} \\
Extreme  & \textbf{71.78}  & \textbf{80.98} & \textbf{43.61} & \textbf{73.75} & \textbf{80.73} &  & \uline{60.74}  & 66.26          & \textbf{37.14} & 62.15          & 68.06          &  & 51.53            & \uline{71.78}  & \uline{17.08}  & \uline{53.53}  & \uline{72.08}  \\
Central  & 66.87           & 72.39          & 33.55          & 66.04          & 73.64          &  & 52.76          & 65.03          & 23.78          & 54.57          & 65.71          &  & 62.58            & 82.21          & 26.33          & 66.45          & 79.84          \\
Scenario & WizardCoder-15B &                &                &                &                &  & Mistral-7B     &                &                &                &                &  & CodeQWen-7B      &                &                &                &                \\
         & Top@5           & Top@10         & Pass@1         & Pass@5         & Pass@10        &  & Top@5          & Top@10         & Pass@1         & Pass@5         & Pass@10        &  & Top@5            & Top@10         & Pass@1         & Pass@5         & Pass@10        \\
Original & 42.33           & 60.12          & 13.40          & 42.92          & 58.44          &  & 44.17          & 52.76          & 19.82          & 43.52          & 52.60          &  & 85.28            & 89.57          & 53.14          & 83.82          & 89.47          \\
IID      & \textbf{63.80}  & \textbf{76.07} & \textbf{31.52} & \textbf{67.65} & \textbf{77.83} &  & 50.92          & 61.35          & 23.08          & \uline{52.95}  & 62.13          &  & 84.05            & \uline{89.57}  & 56.44          & 84.71          & 90.06          \\
Mild     & \uline{55.21}   & 67.48          & 21.42          & 54.72          & \uline{67.82}  &  & \textbf{53.99} & \textbf{65.64} & \uline{26.00}  & \textbf{56.26} & \textbf{67.03} &  & 80.98            & 86.50          & 54.18          & 82.81          & 87.48          \\
Medium   & 53.99           & \uline{68.10}  & 22.79          & 56.03          & 67.15          &  & 48.47          & 62.58          & 25.48          & 52.44          & \uline{64.41}  &  & \textbf{85.89}   & \uline{89.57}  & \uline{59.13}  & \uline{85.65}  & \uline{90.37}  \\
Extreme  & 52.15           & \uline{68.10}  & \uline{24.07}  & \uline{56.67}  & 66.31          &  & \uline{52.76}  & \uline{65.03}  & \textbf{27.61} & 52.76          & 63.26          &  & \uline{85.28}    & \textbf{91.41} & \textbf{61.49} & \textbf{86.66} & \textbf{91.54} \\
Central  & 53.99           & 63.80          & 16.80          & 48.10          & 60.93          &  & 42.94          & 53.37          & 14.20          & 40.06          & 52.23          &  & 90.80            & 92.64          & 59.37          & 88.65          & 92.98          
\end{tblr}
\label{cs}
\end{table*}

\begin{table*}[htbp]
\fontsize{6.5}{6.5}\selectfont
\centering
\caption{Experimental results of different code complexity distributions on EvalRepair-Java.}
\begin{tblr}{
  width = \linewidth,
  colspec = {Q[54]Q[52]Q[54]Q[54]Q[54]Q[56]Q[27]Q[52]Q[54]Q[54]Q[54]Q[56]Q[27]Q[52]Q[54]Q[54]Q[54]Q[56]},
  cells = {c},
  cell{1}{1} = {r=2}{},
  cell{1}{2} = {c=5}{0.27\linewidth},
  cell{1}{8} = {c=5}{0.27\linewidth},
  cell{1}{14} = {c=5}{0.27\linewidth},
  cell{9}{1} = {r=2}{},
  cell{9}{2} = {c=5}{0.27\linewidth},
  cell{9}{8} = {c=5}{0.27\linewidth},
  cell{9}{14} = {c=5}{0.27\linewidth},
  hline{1,9,17} = {-}{},
  hline{3,11} = {2-6,8-12,14-18}{},
  hline{4,8,12,16} = {1-6,8-12,14-18}{},
  colsep = {2pt}
}
Scenario & CodeLlama-13B  &                &                &                &                &  & CodeLlama-7B   &                &                &                &                &  & DeepSeekCoder-7B &                &                &                &                \\
         & Top@5          & Top@10         & Pass@1         & Pass@5         & Pass@10        &  & Top@5          & Top@10         & Pass@1         & Pass@5         & Pass@10        &  & Top@5            & Top@10         & Pass@1         & Pass@5         & Pass@10        \\
Original & 56.44          & 66.87          & 20.76          & 52.36          & 65.59          &  & 53.37          & 64.42          & 25.72          & 54.31          & 63.44          &  & 39.26            & 53.37          & 10.10          & 34.85          & 50.62          \\
IID      & \uline{72.39}  & \uline{78.53}  & 38.84          & \uline{70.41}  & \uline{78.30}  &  & 60.12          & \uline{73.01}  & \uline{30.34}  & 62.75          & \uline{72.77}  &  & 53.37            & \uline{68.10}  & 16.99          & 51.57          & 66.68          \\
Mild     & 69.33          & 76.07          & \uline{40.25}  & 70.06          & 76.91          &  & 53.37          & 66.26          & 27.42          & 57.75          & 67.69          &  & \textbf{63.80}   & \textbf{76.07} & \textbf{26.24} & \textbf{62.47} & \textbf{75.42} \\
Medium   & \textbf{73.01} & \textbf{80.98} & \textbf{43.51} & \textbf{73.22} & \textbf{79.50} &  & \textbf{63.80} & \textbf{75.46} & \textbf{33.51} & \textbf{64.94} & \textbf{75.04} &  & \uline{53.99}    & \uline{68.10}  & \uline{17.37}  & \uline{52.95}  & \uline{69.32}  \\
Extreme  & 68.71          & 76.07          & 37.33          & 69.42          & 76.80          &  & \uline{60.74}  & 69.94          & \uline{30.34}  & \uline{63.27}  & 71.36          &  & 46.63            & 67.48          & 17.08          & 52.30          & 68.76          \\
Central  & 66.87          & 72.39          & 33.55          & 66.04          & 73.64          &  & 52.76          & 65.03          & 23.78          & 54.57          & 65.71          &  & 62.58            & 82.21          & 26.33          & 66.45          & 79.84          \\
Scenario & WizardCoder-15B &                &                &                &                &  & Mistral-7B      &                &                &                &                &  & CodeQWen-7B       &                &                &                &                \\
         & Top@5          & Top@10         & Pass@1         & Pass@5         & Pass@10        &  & Top@5          & Top@10         & Pass@1         & Pass@5         & Pass@10        &  & Top@5            & Top@10         & Pass@1         & Pass@5         & Pass@10        \\
Original & 42.33          & 60.12          & 13.40          & 42.92          & 58.44          &  & 44.17          & 52.76          & 19.82          & 43.52          & 52.60          &  & 85.28            & 89.57          & 53.14          & 83.82          & 89.47          \\
IID      & \textbf{61.35} & \textbf{71.17} & \textbf{26.05} & \textbf{60.23} & \textbf{69.98} &  & 47.85          & \uline{64.42}  & 21.05          & 51.84          & 64.78          &  & 84.66            & \uline{90.80}  & 56.77          & 85.67          & \uline{90.40}  \\
Mild     & \uline{55.21}  & 63.80          & \uline{21.19}  & \uline{52.50}  & 63.16          &  & \uline{53.99}  & \uline{64.42}  & \textbf{26.38} & \uline{53.74}  & 63.47          &  & 82.21            & \uline{90.80}  & 55.40          & 84.94          & 90.15          \\
Medium   & 49.08          & 64.42          & 17.27          & 49.76          & 64.35          &  & 52.76          & \textbf{67.48} & \uline{24.87}  & \textbf{53.98} & \textbf{68.01} &  & \uline{86.50}    & \textbf{92.02} & \uline{56.91}  & \uline{85.88}  & \textbf{91.25} \\
Extreme  & 51.53          & \uline{67.48}  & 17.27          & 50.67          & \uline{65.64}  &  & \textbf{54.60} & \uline{64.42}  & 23.69          & 53.59          & \uline{65.97}  &  & \textbf{88.96}   & 89.57          & \textbf{59.70} & \textbf{86.59} & 90.24          \\
Central  & 53.99          & 63.80          & 16.80          & 48.10          & 60.93          &  & 42.94          & 53.37          & 14.20          & 40.06          & 52.23          &  & 90.80            & 92.64          & 59.37          & 88.65          & 92.98          
\end{tblr}
\label{cc}
\end{table*}

\begin{table*}[htbp]
\fontsize{6.5}{6.5}\selectfont
\centering
\caption{Experimental results of different code embedding distributions on EvalRepair-Java.}
\begin{tblr}{
  width = \linewidth,
  colspec = {Q[54]Q[52]Q[54]Q[54]Q[54]Q[56]Q[27]Q[52]Q[54]Q[54]Q[54]Q[56]Q[27]Q[52]Q[54]Q[54]Q[54]Q[56]},
  cells = {c},
  cell{1}{1} = {r=2}{},
  cell{1}{2} = {c=5}{0.27\linewidth},
  cell{1}{8} = {c=5}{0.27\linewidth},
  cell{1}{14} = {c=5}{0.27\linewidth},
  cell{9}{1} = {r=2}{},
  cell{9}{2} = {c=5}{0.27\linewidth},
  cell{9}{8} = {c=5}{0.27\linewidth},
  cell{9}{14} = {c=5}{0.27\linewidth},
  hline{1,9,17} = {-}{},
  hline{3,11} = {2-6,8-12,14-18}{},
  hline{4,8,12,16} = {1-6,8-12,14-18}{},
  colsep = {2pt}
}
Scenario & CodeLlama-13B  &                &                &                &                &  & CodeLlama-7B   &                &                &                &                &  & DeepSeekCoder-7B &                &                &                &                \\
         & Top@5          & Top@10         & Pass@1         & Pass@5         & Pass@10        &  & Top@5          & Top@10         & Pass@1         & Pass@5         & Pass@10        &  & Top@5            & Top@10         & Pass@1         & Pass@5         & Pass@10        \\
Original & 56.44          & 66.87          & 20.76          & 52.36          & 65.59          &  & 53.37          & 64.42          & 25.72          & 54.31          & 63.44          &  & 39.26            & 53.37          & 10.10          & 34.85          & 50.62          \\
IID      & \textbf{75.46} & \textbf{80.37} & 38.89          & \uline{72.73}  & \textbf{82.28} &  & \textbf{61.96} & \textbf{69.94} & 27.74          & 61.46          & \textbf{71.09} &  & 36.81            & 55.83          & 11.94          & 42.90          & 62.02          \\
Mild     & 69.33          & 77.91          & 40.16          & 68.50          & 76.08          &  & 60.12          & 68.10          & 30.39          & 60.35          & 67.62          &  & 30.67            & 47.85          & 10.05          & 37.00          & 55.77          \\
Medium   & 70.55          & \uline{79.14}  & \uline{42.24}  & 72.54          & 79.85          &  & 61.35          & 68.10          & \textbf{33.03} & 61.84          & 68.98          &  & 50.92            & 68.10          & 16.28          & 50.98          & 68.59          \\
Extreme  & \uline{71.17}  & 77.30          & \textbf{43.98} & \textbf{72.84} & \uline{79.93}  &  & 60.12          & 69.33          & 32.42          & \textbf{62.14} & 69.89          &  & \textbf{55.83}   & \textbf{74.23} & \textbf{17.56} & \textbf{54.42} & \textbf{71.44} \\
Central  & 66.87          & 72.39          & 33.55          & 66.04          & 73.64          &  & 52.76          & 65.03          & 23.78          & 54.57          & 65.71          &  & 62.58            & 82.21          & 26.33          & 66.45          & 79.84          \\
Scenario & WizardCoder-15B &                &                &                &                &  & Mistral-7B      &                &                &                &                &  & CodeQWen-7B       &                &                &                &                \\
         & Top@5          & Top@10         & Pass@1         & Pass@5         & Pass@10        &  & Top@5          & Top@10         & Pass@1         & Pass@5         & Pass@10        &  & Top@5            & Top@10         & Pass@1         & Pass@5         & Pass@10        \\
Original & 42.33          & 60.12          & 13.40          & 42.92          & 58.44          &  & 44.17          & 52.76          & 19.82          & 43.52          & 52.60          &  & 85.28            & 89.57          & 53.14          & 83.82          & 89.47          \\
IID      & \textbf{66.26} & \textbf{73.01} & \textbf{27.98} & \textbf{63.98} & \textbf{74.42} &  & \uline{51.53}  & \uline{66.87}  & 23.12          & \uline{52.98}  & \uline{65.60}  &  & \textbf{85.28}   & \textbf{90.18} & \uline{58.47}  & \uline{85.34}  & \textbf{90.45} \\
Mild     & 55.21          & 65.64          & 24.26          & \uline{57.66}  & \uline{69.25}  &  & 50.92          & \textbf{67.48} & \uline{23.83}  & \textbf{53.97} & \textbf{66.39} &  & 81.60            & 88.34          & 53.37          & 83.79          & 89.38          \\
Medium   & \uline{59.51}  & \uline{66.26}  & 22.89          & 56.55          & 68.80          &  & \textbf{54.60} & \uline{66.87}  & \textbf{25.01} & 52.88          & 64.35          &  & 84.05            & \uline{89.57}  & 53.80          & 84.43          & 90.38          \\
Extreme  & 57.06          & \uline{66.26}  & \uline{24.73}  & 55.80          & 65.20          &  & 50.92          & 61.96          & 22.84          & 50.85          & 62.16          &  & \uline{84.66}    & \uline{89.57}  & \textbf{61.68} & \textbf{86.04} & \uline{90.40}  \\
Central  & 53.99          & 63.80          & 16.80          & 48.10          & 60.93          &  & 42.94          & 53.37          & 14.20          & 40.06          & 52.23          &  & 90.80            & 92.64          & 59.37          & 88.65          & 92.98          
\end{tblr}
\label{ce}
\end{table*}

For the mild Non-IID coding style scenario, as illustrated in Table \ref{cs}, CodeLlama-7B and Mistral-7B exhibit the best repairing capabilities among all data distributions. DeepSeekCoder-7B achieves the best performance in terms of all metrics with $Top@10$ and $Pass@10$ at 76.07\% and 75.42\%, under the scenario of mild Non-IID code complexity as shown in Table \ref{cc}. On the other hand, we observe that the DeepSeekCoder-7B model is enhanced the most with an increase of 14.73\% and 16.84\% on $Top@10$ and $Pass@10$ compared to the original model, respectively, as presented in Table \ref{cs}. Remarkably, the DeepSeekCoder-7B model is improved by 22.27\% and 24.80\% on $Top@10$ and $Pass@10$, under the mildly heterogeneous code complexity scenario as presented in Table \ref{cc}. The fine-tuned Mistral-7B shows the largest improvement under the scenario of mildly heterogeneous code embedding in Table \ref{ce}, with an increase of 14.72\% and 13.79\% on $Top@10$ and $Pass@10$. The results demonstrate that the federated fine-tuning approach is effective in enhancing the performance of LLMs for program repair under the mild degree of heterogeneity so far.

As the heterogeneity increases, we further investigate whether fine-tuning under the medium Non-IID scenario can enhance the repairing ability of LLMs. The results in Table \ref{cs} and Table \ref{cc} show that DeepSeekCoder-7B, CodeLlama-13B and CodeLlama-7B outperform the other data distributions under the medium Non-IID scenario, particularly with the DeepSeekCoder-7B model under the heterogeneous coding style scenario, achieving the largest increase of 22.09\% and 25.62\% on $Top@10$ and $Pass@10$ compared to the original model. We also notice from Table \ref{ce} that CodeLlama-7B experiences the smallest improvement of the bug fixing capability of 3.68\% and 5.54\% in terms of $Top@10$ and $Pass@10$. The results indicate that all of the LLMs can be enhanced except for CodeQWen-7B, which will be discussed in subsequent analysis, under the scenario of medium data heterogeneity.

In the extreme Non-IID scenario, each client only holds a single category of data. Different from previous findings that indicate the severity of Non-IID data correlates with increased impact on the performance \cite{li2022federated,luo2021no,hsieh2020non}, our study reveals that fine-tuning LLMs under even the most heterogeneous Non-IID scenario is able to enhance the program repair performance. For example, the CodeLlama-13B model in Table \ref{cs} achieves the $Top@10$ and $Pass@10$ at 80.98\% and 80.73\% and the DeepSeekCoder-7B model in Table \ref{ce} achieves the $Top@10$ and $Pass@10$ at 74.23\% and 71.44\%, both outperforming the other data distributions across all metrics under the extreme Non-IID scenario. We also notice that the DeepSeekCoder-7B model fine-tuned under the extreme scenario obtains the largest increment on both $Top@10$ and $Pass@10$ compared to other LLMs. For instance, Table \ref{cs} presents that the $Top@10$ and $Pass@10$ of the fine-tuned DeepSeekCoder-7B are increased by 18.41\% and 21.46\%. The results further demonstrate that heterogeneous code can benefit the fine-tuning of LLMs for program repair.

Similarly, it is also evident that fine-tuning LLMs under Non-IID scenarios can achieve better performance than centralized fine-tuning. We observe that except for DeepSeekCoder-7B and CodeQWen-7B, all of the other LLMs outperform the central approach across all data distributions, especially on $Top@10$ and $Pass@10$. For instance, as Table \ref{cc} presents, the Mistral-7B model under medium Non-IID scenario achieves an increase of 14.11\% and 13.92\% in terms of $Top@10$ and $Pass@10$ compared to the central approach. An increase of 13.36\% on $Pass@1$ can also be observed on CodeLlama-7B under the extreme Non-IID scenario in Table \ref{cs}.

In order to further confirm the effectiveness of federated fine-tuning under different data distributions, we also conduct statistical tests to validate whether and to what extent it can improve the repairing capability of LLMs. Table \ref{p} reveals the significance of the improvement achieved by fine-tuning with different data distributions. Specifically, the results in Table \ref{p} present the p-values and effect sizes of the tests between fine-tuning with all distributions and the original model, as well as between different Non-IID distributions and the IID distribution, across the LLMs for each code feature. 

\begin{table*}
\fontsize{6.5}{6.5}\selectfont
\centering
\caption{Experimental results of Wilcoxon Signed-Rank Test and Cliff's Delta Measurement across different data distributions}
\begin{tblr}{
  width = \linewidth,
  colspec = {Q[87]Q[65]Q[52]Q[65]Q[52]Q[69]Q[54]Q[67]Q[54]Q[65]Q[52]Q[60]Q[52]Q[60]Q[69]},
  cells = {c},
  cell{1}{1} = {r=2}{},
  cell{1}{2} = {c=2}{0.117\linewidth},
  cell{1}{4} = {c=2}{0.117\linewidth},
  cell{1}{6} = {c=2}{0.123\linewidth},
  cell{1}{8} = {c=2}{0.121\linewidth},
  cell{1}{10} = {c=2}{0.117\linewidth},
  cell{1}{12} = {c=2}{0.111\linewidth},
  cell{1}{14} = {c=2}{0.129\linewidth},
  hline{1,6-7} = {-}{},
  hline{3} = {2-15}{},
  colsep = {2pt}
}
Code Feature       & IID \& Original &                & Mild \& Original &                & Medium \& Original &                & Extreme \& Original &                & Mild \& IID &                & Medium \& IID &                & Extreme \& IID &                \\
                   & p-value       & {Effect\\Size} & p-value        & {Effect\\Size} & p-value          & {Effect\\Size} & p-value           & {Effect\\Size} & p-value   & {Effect\\Size} & p-value     & {Effect\\Size} & p-value      & {Effect\\Size} \\
{Coding\\Style}    & 5.25E-06      & 0.29           & 2.05E-07       & 0.34           & 2.56E-06         & 0.36           & 2.56E-09          & 0.34           & 0.37      & 0.01           & 0.14        & 0.05           & 0.03         & 0.06           \\
{Code\\Complexity} & 3.73E-09      & 0.34           & 5.32E-06       & 0.35           & 1.86E-09         & 0.33           & 2.56E-06          & 0.31           & 0.58      & -0.03          & 0.07        & 0.02           & 0.16         & -0.06          \\
{Code\\Embedding}  & 4.80E-06      & 0.31           & 5.59E-05       & 0.24           & 3.90E-06         & 0.34           & 2.85E-06          & 0.35           & 9.22E-06  & -0.11          & 0.58        & -0.04          & 0.66         & -0.02          \\
Overall            & 1.41E-15      & 0.31           & 1.07E-13       & 0.31           & 4.74E-16         & 0.35           & 5.65E-16          & 0.33           & 0.05      & -0.04          & 0.12        & 0.01           & 0.56         & -2.72E-03      
\end{tblr}
\label{p}
\end{table*}

We observe that all of the differences between the performance of federated fine-tuning with different distributions and the original models (i.e., from \textit{IID \& Original} to \textit{Extreme \& Original}) are statistically significant (i.e., p-value<0.05). For instance, the LLMs fine-tuned under the scenario of extremely heterogeneous coding style yield a p-value of $2.56e^{-6}$ and an effect size of 0.34, indicating that the difference is statistically significant while the difference is also a positive improvement. These results demonstrate that the LLMs fine-tuned under various Non-IID scenarios, including even the extreme one, are confirmed to facilitate significant improvement. We also notice that all effect sizes except for the scenario of mildly heterogeneous code embedding, which yields an effect size of 0.24, fall into the range of medium effect size (i.e., $[0.28,0.43)$), further validating that federated fine-tuning with either IID or Non-IID distributions leads to a promising increase in the repairing capability of LLMs across the evaluation metrics.

\begin{tcolorbox}
\textbf{Finding 4:} Federated fine-tuning with different data distributions, including the most extreme Non-IID scenario, is able to improve the repairing capabilities as significantly as the IID scenario, providing the insight that real-world industries can benefit from collaborative learning regardless of diverse data distributions.
\end{tcolorbox}

On the other hand, contrary to most previous studies regarding federated learning on traditional DL tasks where Non-IID data mostly results in significant performance degradation compared to the ideal IID distribution \cite{kairouz2021advances,li2020federated,li2022federated}, we find that fine-tuning LLMs with diverse Non-IID distributions is overall negligibly affected and in many cases even outperforming the IID scenario. 

Firstly, we observe that fine-tuning under the Non-IID scenarios on specific LLMs exhibits a slight decline in performance compared to the IID scenario. For example, as presented in Table \ref{cs}, we notice a maximal decrease of 6.14\% for CodeLlama-13B on $Top@10$ in terms of fine-tuning with medium Non-IID distribution and a decrease of 2.46\% on $Pass@10$ under the mild Non-IID scenario, compared to the IID distribution. In Table \ref{cc}, we see that both CodeLlama-13B and CodeLlama-7B are slightly influenced by fine-tuning under some of the heterogeneous code complexity scenarios. There is a maximal decrease of only 3.68\% on $Top@5$ while fine-tuning CodeLlama-13B under the extreme Non-IID scenario and a reduction of 6.75\% on $Top@10$ while fine-tuning CodeLlama-7B with mild Non-IID distribution. The results in Table \ref{ce} show that the LLMs are more susceptible to heterogeneous distributions of code embedding, where all of the LLMs except DeepSeekCoder-7B exhibit slight declines in the performance across the three Non-IID scenarios.

However, the results also reveal that fine-tuning under Non-IID scenarios can outperform the IID scenario in many cases. In particular, Table \ref{cs} presents that the CodeLlama-7B model and DeepSeekCoder-7B model outperform the IID scenario under all Non-IID scenarios. Notably, in the best-performing scenario of DeepSeekCoder-7B (i.e., Medium), the model achieves a significant increase of 15.34\% and 18.68\% compared to the IID distribution. Mistral-7B and DeepSeekCoder-7B in the heterogeneous code complexity scenarios, as illustrated by Table \ref{cc}, achieve the best repairing performance with all Non-IID distributions. The DeepSeekCoder-7B also achieves the largest increase at 19.02\% and 18.40\% on $Top@5$ and $Top@10$ under the extreme Non-IID scenario compared to the IID scenario from Table \ref{ce}. These results indicate that while the LLMs can be slightly affected by some of the heterogeneous distributions, they still exhibit superior performance compared to the IID scenario in many cases.

To further confirm whether these effects are significant and to determine if they are overall positive or negative effects, as they have specific pros and cons over different scenarios, we also conduct significance tests between the Non-IID scenarios and the IID scenario. The p-values and effect sizes of the differences between the Non-IID distributions and the IID (i.e., from \textit{Mild \& IID} to \textit{Extreme \& IID}) are presented in Table \ref{p}.

Specifically, the results in Table \ref{p} show that most of the differences between Non-IID scenarios and the IID scenario are not significant. The only exceptions are that the difference between the mildly heterogeneous code embedding scenario and the IID scenario yields a p-value of $9.22e^{-6}$ and the difference between the extremely heterogeneous coding style scenario and the IID scenario yields a p-value of 0.03. However, we notice their corresponding values of Cliff's delta are -0.11 and 0.06, which indicate negligible effects. Therefore, even the most significant differences have minimal practical impact on the repairing performance of the federated fine-tuning approach. In addition, the overall results across the three code features also yield p-values$\geq$0.05 with trivial effect sizes, further confirming that fine-tuning with heterogeneous code has negligible impact on the performance of bug fixing.

\begin{tcolorbox}
\textbf{Finding 5:} Different degrees of data heterogeneity from mild Non-IID to extreme Non-IID have negligible impact on the performance of LLM fine-tuning compared to IID data distribution, indicating that it is feasible for real-world industries holding datasets with either similar or distinct data distributions to collaborate in the context of LLM fine-tuning in federated learning.
\end{tcolorbox}

Despite the fact that the overall impact of heterogeneous code is negligible, we find that WizardCoder-15B only achieves the best performance with the IID data distribution across all code features. While the WizardCoder-15 model can also benefit from fine-tuning with heterogeneous code, we notice that compared to the IID scenario, the Non-IID scenarios consistently cause degradation on the performance of WizardCoder-15B, particularly for the extremely heterogeneous coding style scenario where the $Pass@10$ decreases by up to 11.52\%, as presented in Table \ref{cs}. The results in Table \ref{cc} and Table \ref{ce} also demonstrate that WizardCoder-15B experiences a decline from over 70\% to over 60\% across the Non-IID scenarios, especially on $Top@10$ and $Pass@10$. These results indicate that WizardCoder-15B is more susceptible to heterogeneous code compared to the other LLMs. 

It is also worth noting that CodeQWen-7B exhibits stable and consistent performance across all code features and data distributions. For example, we observe that CodeQWen-7B achieves a maximal improvement of 2.07\% on $Pass@10$ under the extreme Non-IID scenario as presented by Table \ref{cs}, and the greatest increase of 2.45\% on $Top@10$ is achieved by fine-tuning with the medium Non-IID distribution as illustrated in Table \ref{cc}. On the other hand, CodeQWen-7B is shown to be hardly affected by heterogeneous code compared to the IID distribution. In particular, we see that regardless of the increase or decrease in performance, the bug fixing capability of CodeQwen on the $top@10$ and $Pass@10$ metrics remains around 90\% across all data distributions and code features, demonstrating remarkable stability.

\begin{tcolorbox}
\textbf{Finding 6:} Despite that the bug fixing capability of WizardCoder-15B can be enhanced through fine-tuning with various data distributions, it remains more susceptible to heterogeneous code compared to other models. In contrast, CodeQWen-7B demonstrates remarkable stability and consistency across all code features and data distributions, with minimal performance fluctuation compared to other LLMs.
\end{tcolorbox}

\subsection{RQ3: Impact of federated algorithms}

\noindent\textbf{[Experiment Goal for RQ3]:} We investigate the impact of applying various types of federated algorithms optimized for different phases of federated learning on fine-tuning LLMs for program repair.

\noindent\textbf{[Experiment Design for RQ3]:} As described in the federated learning process in section \ref{ff}, local fine-tuning and federated aggregation are the key components of federated learning. On the client side, each local device updates its model based on the local data towards a local objective, while on the server side, the updated parameters are further aggregated into a new global model according to a specific strategy. We employ various algorithms to comprehensively explore different types of federated algorithms based on the two sides. In addition, we extend our investigation to personalized learning, which is a different learning paradigm that takes into account the individual characteristics and preferences of each client.

\textbf{Client-side Optimization.} In the context of federated learning, client-side optimization plays a crucial role in enhancing the efficiency and effectiveness of local model training on participating devices. To establish a baseline for client-side optimization, we employ FedProx \cite{li2020federated}, a widely used federated algorithm specifically designed to optimize the local training process. FedProx introduces a proximal term in the local objective function, which serves to improve the quality of local models that may not have undergone sufficient training. We evaluate FedProx to investigate the role of client-side optimization during the federated fine-tuning process.

\textbf{Server-side Optimization.} Server-side optimization is crucial for aggregating updates from diverse client models into a stable global model. Therefore, we leverage FedSWA \cite{caldarola2022improving} as another baseline since it applies a stochastic weight averaging strategy for federated aggregation on the server side to promote better generalizability in federated learning. FedSWA enhances generalization by encouraging convergence toward flatter minima, which is beneficial for distributed data across clients. Extensive testing on various tasks validates its robustness across different applications. We evaluate it in the LLM-based federated fine-tuning setting to further investigate its effectiveness.

\textbf{Optimization on Both Sides.} We also evaluate the performance of FedOPT \cite{reddi2021adaptive}, which performs adaptive optimization on both the client side and server side. FedOPT employs adaptive optimization methods on the server and clients for faster convergence compared to traditional methods, robust to noisy gradients, and provides theoretical convergence guarantees that ensure reliable performance across scenarios. We evaluate FedOPT as another type of federated algorithm to assess its performance in comparison to other types of algorithms.

\textbf{Personalized Learning.} Furthermore, we take personalized federated learning into consideration since it has emerged as an alternative federated learning approach to further improve model performance in recent studies \cite{pillutla2022federated,li2021ditto,tang2021personalized,t2020personalized}. The main difference from traditional federated learning is that personalized federated learning is able to handle client-specific characteristics. While federated learning focuses on training a single global model that performs well on average across all clients, personalized federated learning takes a step further by creating personalized models, such as retaining parts of the updated model parameters in each local client \cite{pillutla2022federated,li2021ditto}, that adapt to the unique characteristics of each client. However, it still remains unclear whether LLM fine-tuning can benefit from personalized learning. Therefore, we leverage a typical and widely used algorithm pFedMe \cite{t2020personalized} as the personalization technique to evaluate the performance of LLMs fine-tuned for program repair. pFedMe utilizes Moreau envelopes as regularized loss functions to decouple personalized model updates from global model learning, allowing each client to optimize its personalized model based on local data \cite{t2020personalized}. 

We compare FedAvg \cite{mcmahan2017communication}, the foundational federated learning algorithm that aggregates locally trained models by averaging their parameters as described in Section \ref{ff}, with the other four baselines to investigate the impact of different types of federated algorithms on LLM fine-tuning.

\textbf{Algorithm Ranking.} Given that different algorithms may exhibit varied performance across diverse LLMs and metrics, we use the Borda count method \cite{rothe2019borda}, which is one of the typical and most important voting rules, to evaluate the overall performance of the algorithms. With the comprehensive nature of Borda count in capturing the strength accurately of each candidate \cite{boehmer2023rank}, we aim to obtain the aggregate ranking that best reflects the overall strengths of the algorithms since several metrics are used to assess each algorithm. Initially, each algorithm receives a Borda score based on their rankings across all voters (i.e., metrics). For example, in a selection with $m$ candidates, the top-ranked candidate is assigned $m$ points while the second-ranked gets $m-1$ points and 1 point for the last candidate. By summing these points across all metrics, the Borda count identifies a winner with broad support, rather than simply based on a majority of the number of first-place performance.

\noindent\textbf{[Experimental Results for RQ3]:} Table \ref{rq3} presents the evaluation results of fine-tuning with different federated algorithms. We notice that FedAvg barely achieves the best performance but performs the second best on most of the metrics across all LLMs. As Table \ref{rq3} shows, FedAvg only achieves the highest $Top@10$ of 79.75\% on CodeLlama-13B and the highest $Top@10$ and $Pass@10$ of 71.78\% and 68.54\% on DeepSeekCoder-7B. Other than that, however, FedAvg reaches the second best performance for most of the other cases. For example, FedAvg performs only slightly inferior to FedSWA on WizardCoder-15B, and it achieves the second-best performance on $Top@5$, $Pass@1$, and $Pass@5$ on both CodeLlama-7B and Mistral-7B.

\begin{figure}[htbp]
\centerline{\includegraphics[width=\textwidth]{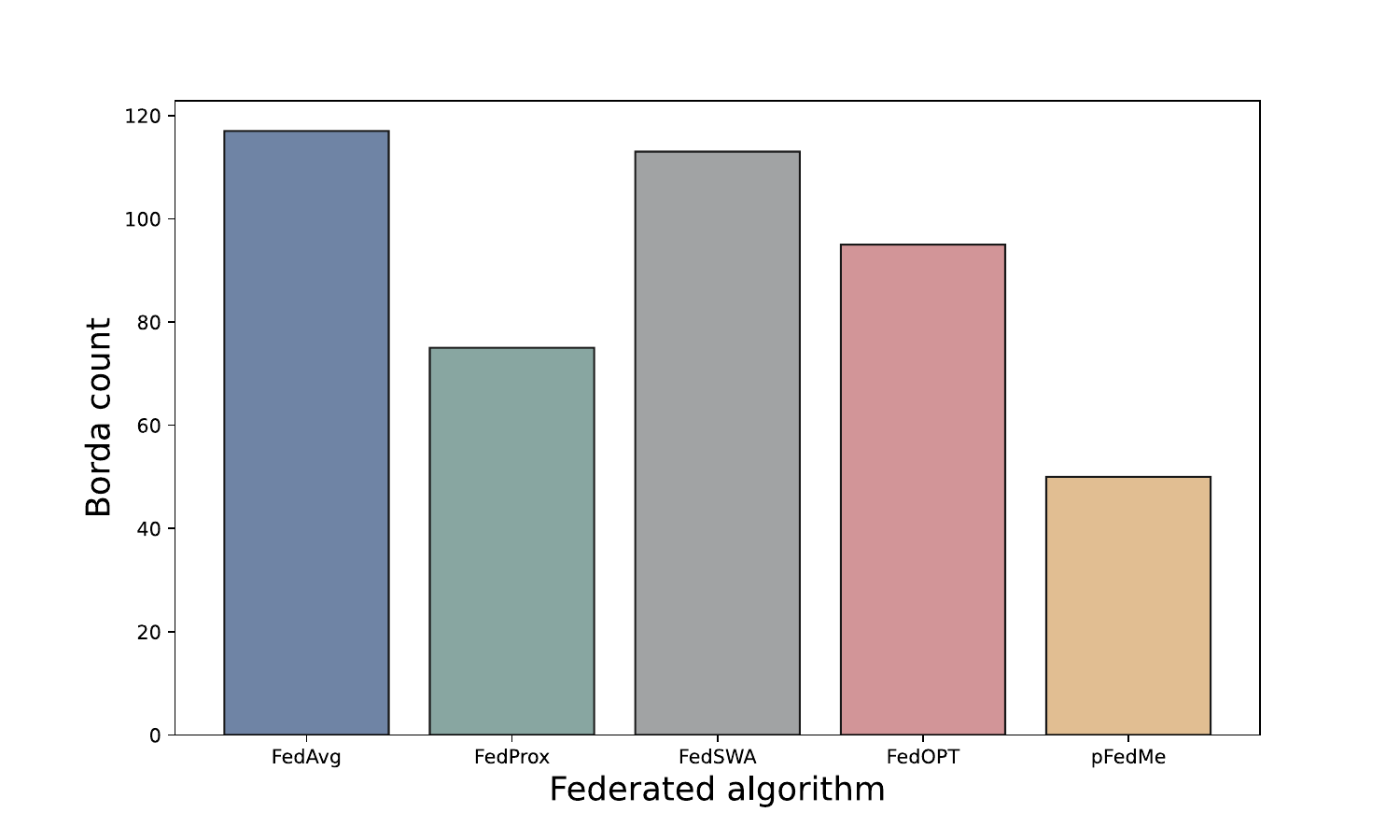}}
\caption{The Borda count of different federated algorithms.}
\label{borda}
\end{figure}

To further validate the effectiveness of the algorithms, we perform the typical Borda count voting algorithm to capture the overall strength of each candidate algorithm. The results of the algorithm ranking is presented in Figure \ref{borda}. The voting results rank the federated algorithms as: \textbf{FedAvg > FedSWA > FedOPT > FedProx > pFedMe}. The results illustrate that FedAvg achieves the highest overall Borda count of 117 points among all algorithms. This is also intuitive since FedAvg consistently achieves the second-best performances under most scenarios while others present varying performances on different LLMs. Therefore, despite that FedAvg does not achieve the best performance across each LLM, the results demonstrate that the overall performance of FedAvg remains promising, demonstrating its effectiveness as the foundational federated algorithm for aggregating knowledge from diverse clients for program repair.

\begin{table*}[htbp]
\fontsize{6.5}{6.5}\selectfont
\centering
\caption{Experimental results of different federated learning algorithms on EvalRepair-Java.}
\begin{tblr}{
  width = \linewidth,
  colspec = {Q[54]Q[52]Q[54]Q[54]Q[54]Q[56]Q[27]Q[52]Q[54]Q[54]Q[54]Q[56]Q[27]Q[52]Q[54]Q[54]Q[54]Q[56]},
  cells = {c},
  cell{1}{1} = {r=2}{},
  cell{1}{2} = {c=5}{0.27\linewidth},
  cell{1}{8} = {c=5}{0.27\linewidth},
  cell{1}{14} = {c=5}{0.27\linewidth},
  cell{8}{1} = {r=2}{},
  cell{8}{2} = {c=5}{0.27\linewidth},
  cell{8}{8} = {c=5}{0.27\linewidth},
  cell{8}{14} = {c=5}{0.27\linewidth},
  hline{1,15} = {-}{},
  hline{3,10} = {2-6,8-12,14-18}{},
  hline{8} = {1-6,8-12,14-18}{},
  colsep = {2pt}
}
Scenario & CodeLlama-13B           &                &                         &                         &                         &  & CodeLlama-7B   &                &                &                &                &  & DeepSeekCoder-7B        &                         &                         &                         &                         \\
         & Top@5                   & Top@10         & Pass@1                  & Pass@5                  & Pass@10                 &  & Top@5          & Top@10         & Pass@1         & Pass@5         & Pass@10        &  & Top@5                   & Top@10                  & Pass@1                  & Pass@5                  & Pass@10                 \\
FedAvg   & 65.64                   & \textbf{79.75} & \uline{31.76}           & \uline{67.49}           & 77.71                   &  & \uline{57.67}  & 66.87          & \uline{28.41}  & \uline{60.35}  & \textbf{70.10} &  & \uline{48.47}           & \textbf{71.78}          & \uline{15.24}           & \uline{49.12}           & \textbf{68.54}          \\
FedProx  & 55.83                   & 63.80          & 19.49                   & 52.28                   & 65.76                   &  & 56.44          & \uline{67.48}  & 21.57          & 52.80          & 65.66          &  & \textbf{50.31}          & \uline{67.48}           & \textbf{16.99}          & \textbf{51.15}          & \uline{68.01}           \\
FedSWA   & \uline{66.26}           & \uline{79.14}  & 30.86                   & 66.55                   & \uline{77.73}           &  & \textbf{58.90} & \textbf{71.17} & \textbf{29.45} & \textbf{60.57} & \uline{70.08}  &  & 42.94                   & 61.35                   & 12.60                   & 43.46                   & 61.60                   \\
FedOPT   & \textbf{\textbf{67.48}} & 75.46          & \textbf{\textbf{36.95}} & \textbf{\textbf{69.15}} & \textbf{\textbf{78.12}} &  & 54.60          & 63.80          & 24.82          & 56.00          & 66.25          &  & 47.85                   & 65.03                   & 14.63                   & 46.93                   & 64.06                   \\
pFedMe   & 58.90                   & 72.39          & 21.85                   & 56.43                   & 70.97                   &  & 53.99          & 63.80          & 26.52          & 55.01          & 63.85          &  & 21.47                   & 42.94                   & 6.61                    & 26.44                   & 42.23                   \\
Scenario & WizardCoder-15B         &                &                         &                         &                         &  & Mistral-7B     &                &                &                &                &  & CodeQWen-7B             &                         &                         &                         &                         \\
         & Top@5                   & Top@10         & Pass@1                  & Pass@5                  & Pass@10                 &  & Top@5          & Top@10         & Pass@1         & Pass@5         & Pass@10        &  & Top@5                   & Top@10                  & Pass@1                  & Pass@5                  & Pass@10                 \\
FedAvg   & \uline{54.60}           & \uline{66.87}  & \uline{22.70}           & \uline{56.35}           & \uline{67.78}           &  & \uline{52.15}  & 65.03          & \uline{22.27}  & \uline{52.39}  & 64.37          &  & \uline{87.73}           & \uline{92.64}           & \uline{58.90}           & 86.65                   & 91.80                   \\
FedProx  & 43.56                   & 58.90          & 15.34                   & 46.20                   & 60.61                   &  & 51.53          & \textbf{66.26} & \textbf{22.89} & \textbf{53.18} & \textbf{65.37} &  & 83.44                   & 90.18                   & 50.92                   & 82.63                   & 89.23                   \\
FedSWA   & \textbf{61.35}          & \textbf{71.17} & \textbf{25.29}          & \textbf{58.70}          & \textbf{69.70}          &  & \textbf{52.76} & \uline{65.64}  & 21.71          & 52.12          & \uline{65.31}  &  & 86.50                   & 92.02                   & 57.57                   & \uline{87.00}           & \uline{92.08}           \\
FedOPT   & 53.99                   & 65.03          & 18.74                   & 51.43                   & 63.37                   &  & 37.42          & 47.24          & 14.87          & 36.33          & 46.45          &  & \textbf{\textbf{88.34}} & \textbf{\textbf{94.48}} & \textbf{\textbf{63.10}} & \textbf{\textbf{88.14}} & \textbf{\textbf{93.50}} \\
pFedMe   & 46.63                   & 63.19          & 14.58                   & 45.17                   & 61.59                   &  & 43.56          & 53.37          & 18.64          & 43.02          & 54.13          &  & 83.44                   & 89.57                   & 53.28                   & 84.40                   & 90.96                   
\end{tblr}
\label{rq3}
\end{table*}

\begin{tcolorbox}
\textbf{Finding 7:} As the foundational and the most widely applied federated algorithm, FedAvg is rarely optimal in the bug fixing ability across all LLMs, but achieves the best overall performance instead, which demonstrates its stability and high practicality in terms of fine-tuning LLMs for program repair in federated learning.
\end{tcolorbox}

At the same time, other algorithms do not exhibit consistency on each LLM and appear to exhibit varying performance across different LLMs. Among other types of federated algorithms we evaluated, FedSWA emerges as the top on CodeLlama-7B and WizardCoder-15B, achieving a $Top10$ of 71.17\% on CodeLlama-7B and a $Pass@10$ of 69.70\% on WizardCoder-15B. We also notice that it achieves the second-best performance on certain metrics across LLMs such as CodeLlama-13B, Mistral-7B, and CodeQWen-7B, highlighting the efficacy and potential of server-side optimization in improving the bug fixing capability of the LLMs in federated fine-tuning. In addition, FedSWA achieved an overall Borda count of 113 points, which is only 4 points lower than FedAvg, demonstrating a promising overall repairing ability.

On the other hand, despite FedOPT and FedProx not achieving comparable overall performance to FedAvg and FedSWA as presented in Figure \ref{borda}, they still demonstrate promising results in specific scenarios. FedOPT showcases its superiority on CodeLlama-13B and CodeQWen-7B, outperforming the other algorithms in these cases. In particular, FedOPT achieves the best $Top@10$ and $Pass@10$ of up to 94.48\% and 93.50\% on CodeQWen-7B, which underscores the significance of employing optimization techniques on both the client and server sides for achieving optimal results with these models. Moreover, the results in Table \ref{rq3} also reveal that FedProx demonstrates comparable performance on DeepSeekCoder-7B and Mistral-7B, achieving optimal results on some of the metrics. For example, FedProx achieves the highest $Top@10$ and $Pass@10$ of 66.26\% and 65.37\% on Mistral-7B, which emphasizes the effectiveness of client-side optimization in fine-tuning for certain LLMs. These results suggest that while FedOPT and FedProx may not be the best choice for all scenarios, they can still be valuable alternatives for specific LLMs and use cases.

\begin{tcolorbox}
\textbf{Finding 8:} Different types of federated algorithms excel in optimizing the performance of specific LLMs, highlighting the importance of considering the specific characteristics and requirements of each LLM when selecting the appropriate optimization approach. By tailoring the optimization process to the unique properties of each LLM, the fine-tuning performance can be effectively enhanced for program repair.
\end{tcolorbox}

However, while different types of federated algorithms demonstrate specific strengths in different scenarios, the results show that the personalization algorithm pFedMe consistently underperforms compared to others in our evaluation. According to the overall ranking results in Figure \ref{borda}, pFedMe only achieves 50 points of Borda count. As presented in Table \ref{rq3}, we notice that pFedMe causes devastating performance degradation on DeepSeekCoder-7B where the $Top@10$ and $Pass@10$ decrease by as much as 28.84\% and 26.31\% compared to FedAvg. It also leads to a reduction of 12.89\% and 11.24\% in terms of $Top@10$ and $Pass@10$ compared to FedProx on Mistral-7B. In addition, we also find that CodeQWen-7B still remains the least affected model, with a decrease of 4.91\% and 2.54\% on $Top@10$ and $Pass@10$ compared to FedOPT, aligning with the previously noted consistency of CodeQWen-7B. These observations on the personalization technique indicate that personalized learning for LLM fine-tuning remains a challenge and still requires further investigation.

\begin{tcolorbox}
\textbf{Finding 9:} Despite the potential benefits of personalized learning demonstrated in most traditional federated learning scenarios, the personalized federated algorithm exhibits the worst performance for the fine-tuning of LLMs, indicating that it remains a major challenge to adapt to LLM fine-tuning for program repair in the context of federated learning.
\end{tcolorbox}

%% file: 6.discussion.tex
\section{Discussion}\label{disc}
\textbf{Impact of Dataset Size}. According to the results of RQ1 in Section \ref{ex-rq1}, we found that the local fine-tuning approach, which fine-tunes the LLMs on each client separately without cooperation, showed the worst performance throughout the evaluation, and the performance of centralized fine-tuning that leverages the complete data was surpassed by federated fine-tuning in some cases. The local fine-tuning approach updates the model with only the local dataset within each client, whereas the full dataset is utilized for the centralized fine-tuning approach, suggesting that the size of the dataset plays a crucial role in the fine-tuning process. According to Jiang et al. \cite{jiang2023impact}, the size of a fine-tuning dataset appears to affect the performance of the LLMs. The results of Jiang's study demonstrate that fine-tuning with either too little or too much data can cause degradation in the model performance. While centralized fine-tuning has access to the full dataset and local fine-tuning only involves a fraction of the overall dataset, both fine-tuning methods may suffer from suboptimal performance due to the dataset size. The boundary of such size of the dataset that affects the fine-tuning of LLMs is unclear and remains to be further explored in future works. 

On the other hand, the key point reflected in this issue is that while both local fine-tuning and centralized fine-tuning exhibit instability in fine-tuning the LLMs, federated fine-tuning combines the privacy-preserving benefits of local fine-tuning with the advantages of centralized fine-tuning that fully utilize valuable data, to achieve stable and consistent performance across most scenarios.

\noindent\textbf{Impact of Code Features}. Since we found that different degrees of Non-IID data on each type of code feature could barely affect the performance of federated fine-tuning, it is evident in our study that the performance of the fine-tuned LLMs was not affected across different code features either from an overall perspective. This suggests that companies having code repositories with either different coding styles, code complexities, or code embeddings are able to improve their models by federated fine-tuning under different data distributions, indicating that the benefits of federated fine-tuning have the potential to extend to industries across various domains.

\noindent\textbf{Federated Algorithms}. Our results also suggest that the FedSWA algorithm seems to be particularly well-suited for CodeLlama-based models, as it consistently performed well on CodeLLama-13B, CodeLLama-7B, and WizardCoder-15B, which are all built upon the CodeLlama base model. Future studies could explore a larger variety of models to understand the impact of the fundamental architectures of different LLMs.

\noindent\textbf{Personalized learning}. We also found that personalization did not perform well for LLM fine-tuning in the context of program repair despite their success in other traditional federated learning tasks. This indicates that the unique characteristics of LLMs and the nature of the program repair task may require different approaches. Another difference is that we fine-tune the adapters rather than the original model, whereas in conventional federated learning tasks, the original model with full parameters is processed by personalization techniques. Further research is needed to investigate and develop personalization strategies specifically tailored for LLM fine-tuning on program repair or other code-related tasks to promote client-specific scenarios.


%% file: 7.threats.tex
\section{Threats to Validity}\label{threats}
\textbf{Threats to Internal Validity.} One potential threat to internal validity arises from the code features. The code features used to construct the Non-IID data distributions may not fully capture the heterogeneity that exists in real-world code repositories across different industries. We utilize a private dataset sourced from a company providing online programming services. The variety of code features can be limited since it may not fully represent the diverse code repositories in other industries, such as game development, artificial intelligence, or scientific research, which can produce code repositories that mainly serve different purposes and use totally different fundamental frameworks and libraries. We construct data heterogeneity based on three different code features to mitigate this issue. In future studies, researchers should consider incorporating code repositories from diverse domains to further investigate the impact of heterogeneous collaboration.

\noindent\textbf{Threats to Construct Validity.} While $Pass@k$ provides an unbiased estimation of the model's performance, the choice of $k$ values may introduce a threat to construct validity. We select the values of 5 and 10 for $k$ from the perspective of developers \cite{kochhar2016practitioners,noller2022trust}. Since $Pass@k$ calculates the expected value for the functional correctness, the evaluated results are able to more accurately reflect the true performance of the model as the value of $k$ increases. However, generating more solutions for larger $k$ values on each problem and each model during the inference phase of LLMs is computationally more expensive. To mitigate the bias of the results, we evaluate the fine-tuning methods on six LLMs across more than 1000 problems. Future work can further explore the potential of LLMs on program repair, given sufficient computational resources.


\noindent\textbf{Threats to External Validity.} A significant threat to external validity in APR research, including both traditional and learning-based approaches, is the potential for patch overfitting due to weak test suites. Patches that pass a limited set of test cases may still be incorrect when subjected to more comprehensive testing. Manual verification of generated patches is resource-intensive, particularly for large-scale problem sets. 
To mitigate this issue, we employ the EvalRepair \cite{yang2024multi} benchmark, which substantially enhances and expands the original HumanEval \cite{chen2021evaluating} test suites by incorporating test cases from EvalPlus \cite{liu2024your}. This approach provides a more rigorous evaluation framework, enabling us to identify and filter out potentially faulty patches, thereby enhancing the generalizability and robustness of our fixes.

%% file: 8.conclusion.tex
\section{Conclusion}\label{conclusion}
In this paper, we perform an empirical study on LLM-based federated fine-tuning for program repair, aiming to provide a comprehensive understanding of how federated learning can enhance LLM fine-tuning while preserving data privacy, demonstrating its effectiveness and practicality for both academia and industry. We evaluate federated fine-tuning across six code-related LLMs based on different architectures. In order to align with real-world scenarios, we leverage a private industrial dataset comprising 1239 programming problems to fine-tune the models. The fine-tuned models are further evaluated on an augmented program repair benchmark, which encompasses 163 programming problems and 583 test cases for each problem. 

Our study investigates three key research questions to thoroughly examine the effectiveness and implications of LLM-based federated fine-tuning for program repair: (1) To validate the efficacy of federated fine-tuning for program repair, we compared the performance of federated fine-tuning with standard fine-tuning approaches and the original LLMs. The results demonstrate the competitive bug-fixing capabilities of federated fine-tuning, with federated fine-tuning even outperforming centralized fine-tuning in some cases, highlighting its ability to learn effectively from distributed codes. (2) We explore the impact of heterogeneous code on federated fine-tuning for program repair. We construct different degrees of feature-skewed Non-IID data, from mild scenarios to extreme scenarios, based on various code features, i.e., coding style, code complexity, and code embedding. We compare different distribution scenarios to explore the impact of data heterogeneity on LLM-based fine-tuning. Our findings revealed that data heterogeneity had minimal impact on the performance of LLM fine-tuning in the context of federated learning, contrary to traditional federated learning tasks. This insight suggests that enterprises with diverse code repositories can still benefit from collaborative model training through federated learning without compromising performance. (3) We examine how different phases of optimization in federated learning influence the LLM fine-tuning for program repair. The optimization process in federated learning comprises several important stages, either on the client or server, which is critical to fine-tuning performance. We evaluate federated algorithms that specifically focus on client-side or server-side optimization. In addition, as another effective learning paradigm in federated learning, we also evaluate the effectiveness of personalized learning to further explore the impacts of different types of federated algorithms. Our results show that the widely used FedAvg algorithm demonstrated stability and practicality for LLM fine-tuning, whereas personalized learning algorithms like pFedMe faced challenges in adapting to the unique characteristics of LLMs and the program repair task, indicating the need for further research in developing personalization strategies specifically tailored for LLM fine-tuning.

This study sheds light on the potential of federated learning in facilitating collaborative software development and maintenance while preserving data privacy.